\documentclass[aps,twocolumn,showpacs,showkeys,superscriptaddress,amsmath,amssymb,nofootinbib,floatfix,prc]{revtex4-1}

\usepackage{graphicx}
\usepackage{subfigure}
\usepackage{bm}

\makeatletter

\begin{document}

\title{Transverse-momentum fluctuations in relativistic heavy-ion collisions from event-by-event viscous hydrodynamics}

\author{Piotr Bo\.zek}
\email{Piotr.Bozek@ifj.edu.pl}
\affiliation{The H. Niewodnicza\'nski Institute of Nuclear Physics, Polish Academy of Sciences, PL-31342 Krak\'ow, Poland}
\affiliation{Institute of Physics, Rzesz\'ow University, PL-35959 Rzesz\'ow, Poland}

\author{Wojciech Broniowski}
\email{Wojciech.Broniowski@ifj.edu.pl} 
\affiliation{The H. Niewodnicza\'nski Institute of Nuclear Physics, Polish Academy of Sciences, PL-31342 Krak\'ow, Poland} 
\affiliation{Institute of Physics, Jan Kochanowski University, PL-25406~Kielce, Poland}

\date{8 March 2012}

\begin{abstract}
We analyze event-by-event fluctuations of the transverse momentum in relativistic heavy-ion collisions at 
$\sqrt{s_{NN}}=200$~GeV in the framework based on the fluctuating Glauber-model initial conditions, 
event-by-event ($3+1$)-dimensional viscous hydrodynamics, and statistical hadronization.
We use the scaled fluctuation measure $\langle \Delta p_{Ti} \Delta p_{Tj} \rangle / \langle \langle p_T \rangle \rangle$.
The identified ``geometric'' mechanism of generating 
the transverse-momentum fluctuations from the initial size fluctuations, transmitted to the 
final statistical-hadronization phase with hydrodynamics, is capable of
easily reproducing the magnitude of the effect and
explains the basic features of the data. On the other hand, it is somewhat too strong,
hinting on modification of the popular Glauber approach to the earliest phase of the collision.
We have checked that the considered measure is insensitive of the values of the shear and bulk viscosity 
coefficients, the freeze-out temperature, and the smoothing parameter for the initial distribution.
It remains unaltered in the core-corona picture and is insensitive to the transverse-momentum conservation, 
approximately imposed in the statistical hadronization.
\end{abstract}



\maketitle

\section{Introduction}

In Ref.~\cite{Broniowski:2009fm} a new mechanism for generating the transverse-momentum fluctuations in relativistic heavy-ion collisions
was identified. It is based on the random event-by-event fluctuations of the initial size of the formed system, its subsequent 
hydrodynamic evolution, and statistical hadronization. 
In the present work we further explore and extend this analysis, applying ($3+1$)-dimensional 
[($3+1$)-D)]
viscous event-by-event hydrodynamics. 
The basic idea of Ref.~\cite{Broniowski:2009fm} is as follows: 
Even when we consider a very narrow centrality class of events, {\em e.g.}, with a strictly fixed 
number of wounded nucleons, $N_w$, the size of the initial fireball fluctuates event-by-event due to the random nature 
of the nuclear collision in the Glauber treatment.
These fluctuations are then transferred by hydrodynamics to the fluctuations of the generated transverse flow velocity.
At freeze-out, this translates into the event-by-event fluctuations of the average transverse momentum of hadrons 
produced in the event, $\langle p_T \rangle$. In essence, via simple scaling arguments, 
a more {\em squeezed} initial condition leads to more rapid expansion, larger 
velocity flow, and higher $\langle p_T \rangle$, while a {\em swollen} initial condition leads to slower expansion, 
lower flow, and lower $\langle p_T \rangle$. 
We will now explore this mechanism through the use of state-of-the-art tools, such as {\tt GLISSANDO} \cite{Broniowski:2007nz} Monte Carlo code for the Glauber phase,
$(3+1)$-D event-by-event viscous hydrodynamics \cite{Bozek:2011ua,Bozek:2011if} for the dynamical evolution, and {\tt THERMINATOR} \cite{Kisiel:2005hn,Chojnacki:2011hb} 
for the statistical hadronization at freeze-out.   

The event-by-event $\langle p_T \rangle$ fluctuations in relativistic collisions have been actively studied theoretically
\cite{Gazdzicki:1992ri,Stodolsky:1995ds,Shuryak:1997yj,Mrowczynski:1997kz,Liu:1998xf,%
Voloshin:1999yf,Baym:1999up,Voloshin:2001ei,Korus:2001au,Gavin:2003cb,DiasdeDeus:2003ei,Voloshin:2004th,Mrowczynski:2004cg,AbdelAziz:2005wc,%
Broniowski:2005ae,Prindle:2006zz,Gavin:2006xd,Sharma:2008qr,Mrowczynski:2009wk,Hama:2009pk}
and experimentally
\cite{Adams:2003uw,Adamova:2003pz,Adler:2003xq,Anticic:2003fd,Adams:2004gp,Adams:2005ka,Adams:2005aw,Adams:2005ka,Adams:2006sg,Grebieszkow:2007xz,na49:2008vb,Adamova:2008sx,%
Agakishiev:2011fs}, as they may reveal
relevant details of the dynamics of the system, more accurate than contained in the one-body observables. 
Moreover, they are expected be sensitive to the critical phenomena at the phase transition, providing an important probe 
for these effects.

Throughout the paper we use the notation
\begin{eqnarray}
 \langle . \rangle, \;\;\; \langle \langle . \rangle \rangle \label{eq:backets}
\end{eqnarray}
to indicate averaging in a given event, and averaging of the single-event averages over all events, respectively. 

The structure of the paper is as follows: in Sec.~\ref{sec:init} we give the details of the Monte Carlo simulations of the initial phase, 
focusing on the size fluctuations, Sec.~\ref{sec:hydro} provides some necessary description of the applied $(3+1)$-D viscous 
hydrodynamics, while the statistical hadronization is described in Sec.~\ref{sec:stat}. We then proceed in Sec.~\ref{sec:res} to presenting the 
results, which are compared to the data from the STAR and PHENIX collaborations. We investigate the influence of 
model details on the results of our calculation, finding them very robust. In particular, the STAR measure of the 
event-by-event transverse momentum fluctuations is insensitive to the medium viscosity, freeze-out temperature, or the smoothing parameter 
of the initial distribution of sources. Our final conclusions and discussion is contained in Sec.~\ref{sec:concl}.

\section{Initial state fluctuations in the Glauber approach \label{sec:init}}

The initial condition for hydrodynamics may be obtained 
from the Glauber approach, leading to the successful wounded-nucleon picture  \cite{Bialas:1976ed,Bialas:2008zza} 
(a wounded nucleon is a nucleon that collided inelastically 
at least once) or its descendants, such as the {\em mixed} model \cite{Kharzeev:2000ph,Broniowski:2007nz}. When the initial 
condition is obtained via Glauber Monte Carlo simulations, the distribution of sources (wounded nucleons or positions of binary 
collisions) in the transverse plane {\em fluctuates}, reflecting the randomness in positions of the 
nucleons in the colliding nuclei. This leads to fluctuations of {\em shape}.

The event-by-event fluctuations of the elliptic
component of initial shape have been actively studied, as they lead to significantly enhanced elliptic flow 
\cite{Aguiar:2000hw,Miller:2003kd,Bhalerao:2005mm,Manly:2005zy,Alver:2006wh,Andrade:2006yh,Voloshin:2006gz,Drescher:2006ca,%
Broniowski:2007ft,Hama:2007dq,Voloshin:2007pc,Hama:2009pk,Andrade:2008fa}. 
They also generate odd Fourier components, absent from the event-averaged studies, such as the triangular 
deformation \cite{Alver:2010gr,Alver:2010dn,Petersen:2010cw}, as well as higher-order components of the flow. 
Other interesting phenomena appear as the result of fluctuations, {\em e.g.}, 
the torque effect \cite{Bozek:2010vz} of the reaction planes at forward and backward pseudorapidities, or
the directed flow at central rapidity \cite{Teaney:2010vd,Gardim:2011qn}.

We now describe in some detail the implementation of the Glauber model used in this work. 
The density of charged particles per unit of pseudorapidity, as a function of centrality, can 
be parametrized using a formula \cite{Kharzeev:2000ph,Back:2001xy,Back:2004dy} 
incorporating an admixture of binary collisions, $N_{\rm bin}$, into the wounded-nucleon model in the following way:
\begin{eqnarray}
\frac{d N_{\rm charged}}{d\eta} \propto  \left ( \frac{1-\alpha}{2} N_w+\alpha N_{\rm bin} \right ), \label{eq:mixe}
\end{eqnarray}
where $\alpha$ is a phenomenological parameter, $\alpha=0.145$ for the highest RHIC energy of $\sqrt{s_{NN}}=200$~GeV \cite{Back:2004dy}.
The initial-state simulations are carried out with {\tt GLISSANDO } 
\cite{Broniowski:2007nz}, including a component from binary collisions.
The parameter
$\alpha$ in the initial distribution is somewhat smaller from the value extracted from the final distributions (see the following). 
The difference is due to the longitudinal expansion and entropy production in the $(3+1)$-D viscous hydrodynamic expansion~\cite{Bozek:2011ua}.

The positions of nucleons in each of the colliding
nuclei are randomly generated from a Woods-Saxon distribution, with an additional constraint enforcing the 
short-range repulsion, namely, that the centers of nucleons
in each nucleus cannot be generated closer than the expulsion distance $d=0.9$~fm.
Nucleons from the two colliding nuclei are wounded, or a binary collision occurs, 
when their centers get closer to each other than the distance
$\sqrt{\sigma^{\rm inel}_{NN}/\pi}$, with $\sigma_{NN}^{\rm inel}$ denoting the inelastic nucleon-nucleon cross section. For the highest RHIC
energy of $\sqrt{s_{NN}}=200$~GeV one has $\sigma_{NN}^{\rm inel}=42$~mb.\footnote{One may 
more appropriately use a Gaussian wounding profile instead of the applied hard-sphere wounding profile, 
but the results do not differ significantly in the case of size fluctuations~\cite{Rybczynski:2011wv}.} 

The notion of sources, originally limited to the transverse plane, may be extended on the rapidity dependence of the 
particle emission. Although this extension is not crucial for the present study, focused on the mid-rapidity region, we 
include it for the integrity of the paper.
The spatial pseudorapidity ($\eta_\parallel$) distribution of the emission profile is given as the sum
of contributions from the forward- and backward-moving wounded
nucleons. Within such an extended framework
Bia\l{}as and Czy\.z have properly described \cite{Bialas:2004su} the pseudorapidity distributions of charged particles in 
the $d-Au$ collisions. Therefore, we assume an asymmetric emission profile 
\cite{Bialas:2004su,Bozek:2010bi}
peaked in the forward (backward) rapidity for the forward (backward) moving wounded nucleons, denoted as $f_+(\eta_\parallel)$ ($f_-(\eta_\parallel)$),
\begin{equation}
f_{\pm}(\eta_\parallel)=\left(1\pm \frac{\eta_\parallel}{y_{\rm beam}}\right)f(\eta_\parallel) ,
\label{eq:asy}
\end{equation}
where $y_{\rm beam}$ is the beam rapidity. The initial profile in space-time rapidity is
\begin{equation}
f(\eta_\parallel)=\exp\left(-\frac{(\eta_\parallel-\eta_0)^2}
{2\sigma_\eta^2}\theta(|\eta_\parallel|-\eta_0)
\right) , 
\label{eq:etaprofile}
\end{equation}
with $\eta_0=1.5$, $\sigma_\eta=1.4$ \cite{Bozek:2011ua}.
The initial entropy density is assumed to have a factorized form
\begin{eqnarray}
s(x,y,\eta_\parallel)&=& \kappa \sum_i  f_\pm(\eta_\parallel) g_i(x,y) \left[
(1-\alpha) + N^{coll}_i\alpha \right]. \nonumber \\
\label{eq:em}
\end{eqnarray}
Here $N^{coll}_i$ is the number of collisions of the participant nucleon $i$,
and   
\begin{eqnarray}
   g_i(x,y)= \frac{1}{2\pi w^2} \exp \left[ - \frac{(x-x_i)^2+(y-y_i)^2}{2w^2}\right].
\end{eqnarray}
implements a Gaussian smearing, replacing the point-like source at 
the transverse position $(x_i,y_i)$ with a Gaussian profile.
The smearing parameter is taken to be $w=0.4$~fm, and the overall scale factor is $\kappa=2.5$~GeV.
The parameter  $\alpha$ of the mixed model is fixed to
reproduce the dependence of $dN/d\eta$ on centrality. In the $(3+1)$-D viscous hydrodynamic 
model the optimum value is $\alpha=0.125$ at the top RHIC energies \cite{Bozek:2011ua}.
We remark that the mixed model works also very well 
for the description of multiplicities a the LHC energy of
$\sqrt{s_{NN}}=2.76$~TeV, where $\alpha=0.15$~\cite{Bozek:2010er,Bozek:2011wa}.

\begin{figure}[tb]
\begin{center}
\includegraphics[angle=0,width=0.38 \textwidth]{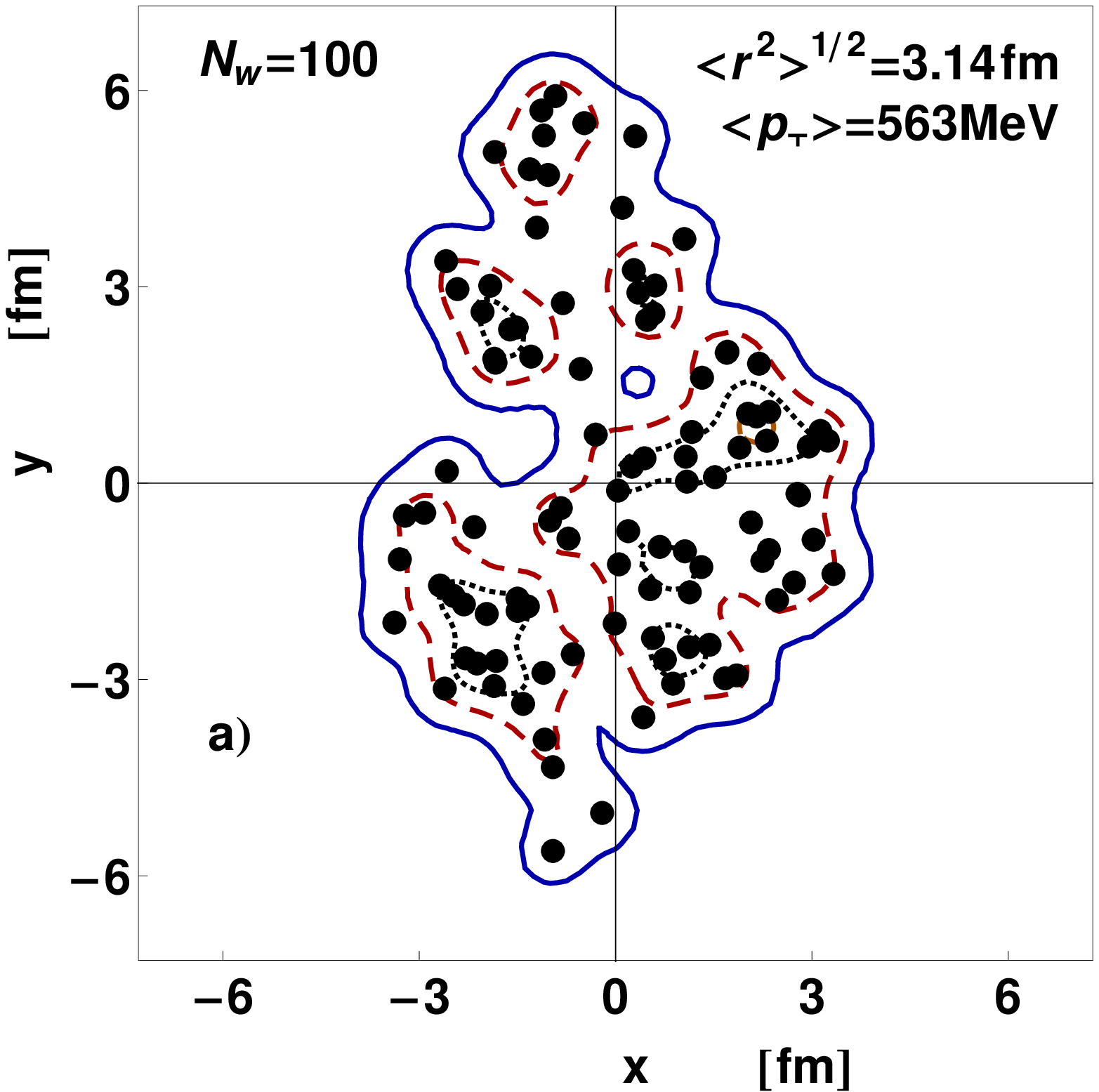} \\
\includegraphics[angle=0,width=0.38 \textwidth]{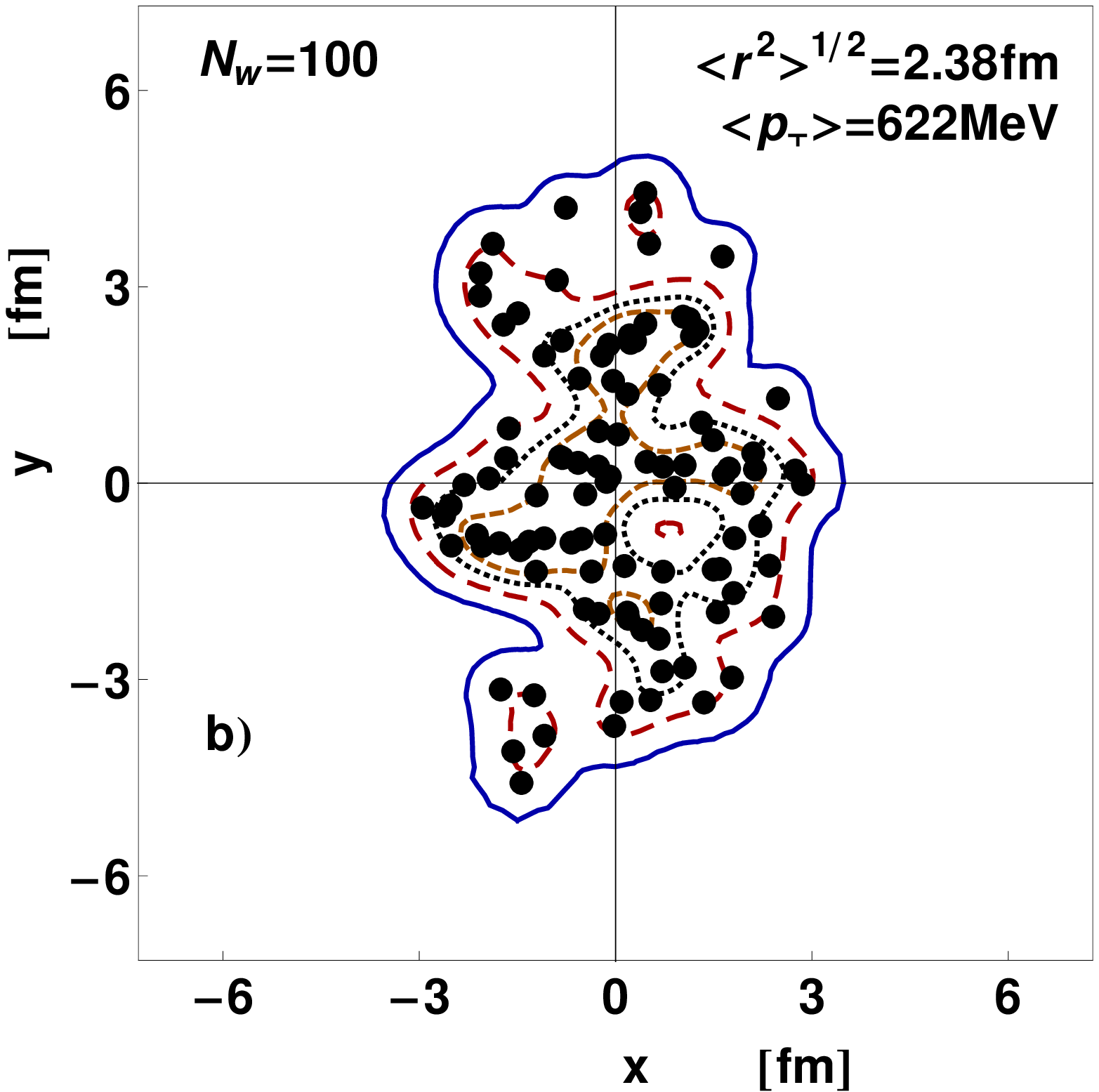} 
\end{center}
\vspace{-7mm}
\caption{(Color online) Two typical configuration of wounded nucleons in the transverse plane (dots) generated 
with {\tt GLISSANDO} and the 
corresponding contours of the smeared density of entropy, $s$.
Solid, dashed, and dotted lines correspond to isentropes at
$s=0.05$, $0.2$, and $0.4$~GeV$^{-3}$, respectively. The densities for the two events have radically different r.m.s. radii of $3.14$ and $2.38$~fm,
respectively, despite the equal number of the wounded nucleons, $N_w=100$.  
\label{fig:photo}} 
\end{figure}

In Fig.~\ref{fig:photo} we show two snapshots of typical configurations of sources in the transverse plane
generated with {\tt GLISSANDO}. The dots indicate the positions of the wounded nucleons. Since we have in mind the distributions as starting conditions for the event-by-event hydrodynamics, we need 
to smear out the point-like distributions. The smearing procedure, although physically motivated and necessary, is 
somewhat arbitrary in introducing a smearing scale.
In Fig.~\ref{fig:photo}, the contours show the smeared entropy density, $s$, with $w=0.4$~fm. Although both
selected events correspond to the same number of wounded nucleons, $N_w=100$, 
they have radically different r.m.s. radii, which after the hydrodynamic
expansion results in different transverse flows.

\begin{figure}[tb]
\begin{center}
\includegraphics[angle=0,width=0.45 \textwidth]{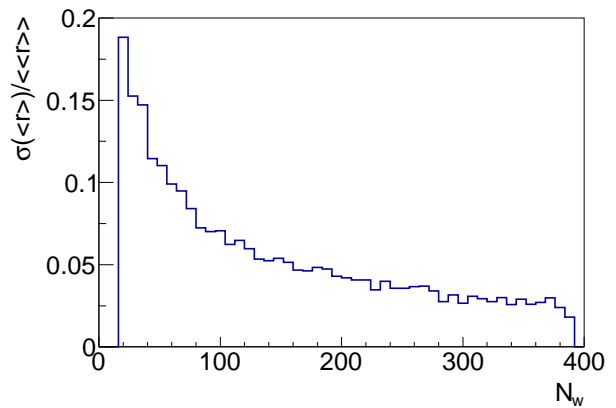} 
\end{center}
\vspace{-7mm}
\caption{(Color online) 
Event-by-event scaled standard deviation of the size parameter $\langle r \rangle$, evaluated at fixed values 
of the number of wounded nucleons $N_w$ from the initial entropy density for events used in 
hydrodynamic simulations.
\label{fig:b}} 
\end{figure}

To have a simple size measure we look at the average transverse size of the initial fireball, defined in each event
via the mean squared radius at the central space-time rapidity%
\begin{eqnarray}
&& \langle r^2 \rangle \equiv \frac{\int dx dy (x^2+y^2) s(x,y,0)}{ \int dx dy \ s(x,y,0)} \ .  \label{eq:def2}
\end{eqnarray}  
In the following we use the notation $\langle r \rangle \equiv \langle r^2 \rangle^{1/2}$. 
The point, clearly seen from Fig.~\ref{fig:photo}, is that even at precisely fixed centrality
the size $\langle r \rangle$ fluctuates \cite{Broniowski:2009fm}. 
The feature is presented quantitatively in Fig.~\ref{fig:b}, 
where we plot the event-by event scaled standard deviation of $\langle r \rangle$ obtained at each 
$N_w$. As expected, $\sigma(\langle r \rangle)/\langle \langle r \rangle \rangle$ is a decreasing function of $N_w$. 

As noted in Ref.~\cite{Broniowski:2009fm}, very similar curves to Fig.~\ref{fig:b} are obtained for other variants of Glauber models, such as models with 
overlaid distributions of particles produced from the sources \cite{Broniowski:2007nz}, simulations 
applying a Gaussian wounding profile \cite{Bialas:2006kw} for the 
$NN$ collisions, or the use of the nucleon distributions including realistic (central) $NN$ 
correlations of Ref.~\cite{Alvioli:2009ab,Broniowski:2010jd,Alvioli:2010yk}.
This means that the behavior of the initial geometry shown in Fig.~\ref{fig:b}
is robust, essentially reflecting the statistical feature of the Glauber approach.

\section{Viscous event-by-event hydrodynamics \label{sec:hydro}}

It is widely believed that a successful and uniform description of the physics of relativistic heavy-ion collisions is 
achieved with the help of relativistic hydrodynamics (for reviews see, e.g., \cite{Kolb:2003dz,Huovinen:2006jp,Florkowski:2010zz}).
{\em Event-by-event} hydrodynamic calculations for fluctuating initial conditions have been performed for perfect fluid
\cite{Andrade:2006yh,Werner:2009fa,Petersen:2010cw,Holopainen:2010gz,Gardim:2011xv,Gardim:2011qn} 
and for the viscous case \cite{Bozek:2011if,Schenke:2010rr,Qiu:2011fi,Chaudhuri:2011pa}, focusing on 
collective flow.

In the second-order {\em viscous} hydrodynamic formalism
\cite{IS,Romatschke:2009im,Teaney:2009qa}, the hydrodynamic equations 
\begin{equation}
\partial_\mu T^{\mu\nu}=0
\end{equation}
with the energy-momentum tensor 
\begin{equation}
T^{\mu\nu}=(\epsilon+p)u^\mu u^\nu-p g^{\mu\nu}+\pi^{\mu\nu}+\Pi \Delta^{\mu\nu}
\label{eq:ideal}
\end{equation}
are supplemented with equations for the stress corrections from the shear,
\begin{equation}
\Delta^{\mu \alpha} \Delta^{\nu \beta} u^\gamma \partial_\gamma 
\pi_{\alpha\beta}=\frac{2\eta \sigma^{\mu\nu}-\pi^{\mu\nu}}{\tau_{\pi}}
-\frac{4}{3}\pi^{\mu\nu}\partial_\alpha u^\alpha,
\label{eq:pidyn}
\end{equation}
and the bulk viscosity, 
\begin{eqnarray}
&& u^\gamma \partial_\gamma \Pi=
\frac{-\zeta \partial_\gamma u^\gamma-\Pi}{\tau_{\Pi}}
-\frac{4}{3}\Pi\partial_\alpha u^\alpha  ,
\label{eq:budyn} \\
&& \sigma_{\mu\nu}=\frac{1}{2}\left( \nabla_\mu  u_\nu
+\nabla_\mu   u_\nu    -\frac{2}{3}\Delta_{\mu    \nu  }\partial_\alpha
u^\alpha\right). \nonumber 
\end{eqnarray}
Here $\nabla^\mu=\Delta^{\mu\nu} \partial_\nu$, while $\eta$ and $\zeta$ denote the shear and bulk 
viscosity coefficients, respectively. In our default calculations we use constant 
$\eta/s=0.08$, $\zeta/s=0.04$ in the hadronic phase, $\tau_\pi={3\eta}/{(T s)}$, and
$\tau_\Pi=\tau_\pi$. 
To test the sensitivity of our results on viscosity, we perform calculations for 
$\eta/s=0.16$, $\zeta/s=0.04$ and $\eta/s=0.08$, $\zeta/s=0.08$ as well. 

The applied equation of state is a {\em crossover} equation of state, interpolating between the lattice-QCD results 
at high temperatures  \cite{Borsanyi:2010cj} and a hadronic gas equation of state at low temperatures.
The construction of the equation of state 
follows the method of Chojnacki and Florkowski \cite{Chojnacki:2007jc} (for details see 
\cite{Bozek:2011ua}).

In this work we apply the event-by-event  ($3+1$)-D viscous hydrodynamics \cite{Schenke:2010rr,Bozek:2011if},
starting the evolution at $0.6$~fm/c. The configurations
of wounded nucleons and binary collisions corresponding to the centrality range $0-70$\%  are generated with
{\tt GLISSANDO}. 
The procedure does not fix the impact parameter for each centrality bin, as the Monte-Carlo scheme picks 
the impact parameter  in each event according to the distribution 
$P(b)=d\sigma_{inel}(b)/(d b \, \sigma_{inel})$ \footnote{This is simply achieved by generating a uniform distribution in $b^2$ 
and accepting those events where at least one $NN$ interaction occurred.}\cite{Broniowski:2001ei}. For each 
configuration of wounded nucleons a hydrodynamic evolution is calculated starting 
from the density (\ref{eq:em}).

\section{Statistical hadronization \label{sec:stat}}

The last stage of our approach is the simulation of the statistical hadronization 
at freeze-out~\cite{Broniowski:2002nf} (for a review, see, e.g. \cite{Florkowski:2010zz}) 
with {\tt THERMINATOR} \cite{Kisiel:2005hn,Chojnacki:2011hb}. The code includes all resonances 
and decay channels from {\tt SHARE} \cite{Torrieri:2004zz}.
The particles (stable and unstable, which subsequently 
decay) are formed at the freeze-out hypersurface according to the Frye-Cooper formula.
In the case of viscous hydrodynamics, the momentum distributions at freeze-out are modified by the 
viscous corrections. The shear and bulk viscosity corrections  are  \cite{Teaney:2003kp}
\begin{equation}
\delta f_{shear}= f_0
\left(1\pm f_0 \right) \frac{1}{2 T^2 (\epsilon+p)}p^\mu p^\nu \pi_{\mu\nu}
\label{eq:dfsh}
\end{equation}
and 
\cite{Gavin:1985ph,*Hosoya:1983xm,*Sasaki:2008fg,Bozek:2009dw},
\begin{equation}
\delta f_{bulk}= C_{bulk}f_0
\left(1\pm f_0 \right)\left(c_s^2 u^\mu p_\mu -\frac{(u^\mu p_\mu)^2-m^2}
{3 u^\mu p_\mu}\right) \Pi   ,
\end{equation}
respectively, with $f_0$ denoting the equilibrium distributions and $c_s$ standing for the velocity of sound.
In the local rest frame the normalization constant is 
\begin{equation}
\frac{1}{C_{bulk}}= \frac{1}{3}\sum_n\int \frac{d^3 p}{(2\pi)^3}\frac{m^2}{E}f_0
\left(1\pm f_0 \right)\left(c_s^2 E -\frac{p^2}{3 E}\right) \ ,
\label{eq:dfbu}
\end{equation} 
where the sum runs over all the hadron species.
The (single-fluid) hydrodynamic evolution uses an equation of state with zero chemical potentials. However, the 
chemical potentials are reintroduced in the Frye-Cooper formula with the ratio $\mu/T$ fixed through the fits
to the particle ratios at the chemical freeze-out, which works properly at the RHIC
energies~\cite{Andronic:2005yp}.

\begin{figure}[tb]
\begin{center}
\includegraphics[angle=0,width=0.47 \textwidth]{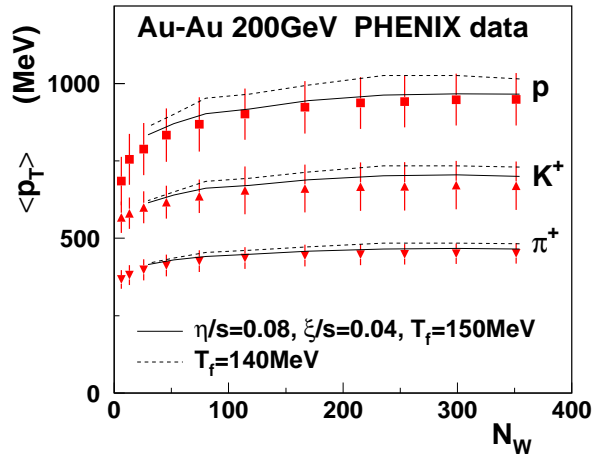} 
\end{center}
\vspace{-7mm}
\caption{(Color online) 
Averaged inclusive transverse momentum vs. number of wounded nucleons, $N_w$. The data (extrapolated to the 
whole $p_T$ range) come from the PHENIX Collaboration~\cite{Adler:2003cb} and show the 
charged pions (down triangle), charged kaons (up triangles), and protons and antiprotons (squares). 
The lines correspond to our model calculation with $\eta/s=0.08$, $\zeta/s=0.04$,
and $T_f=150$~MeV (solid lines), and $T_f=140$~MeV (dashed lines).
\label{fig:pts}} 
\end{figure}  

Before showing the $p_T$-correlation results, let us state that our approach properly describes the relevant one-body features of the collisions, 
in particular, the transverse-momentum spectra. As an example, in Fig.~\ref{fig:pts} we show the inclusive 
average transverse momentum as the function of $N_w$ for pions, kaons, and protons and antiprotons for 
our default parameters $T_f=150$~MeV, $\eta/s=0.08$, $\zeta/s=0.04$ (solid lines). The result compares 
favorably to the PHENIX data \cite{Adler:2003cb}. The agreement is important, as it shows 
that we have the correct one-body background to study correlations.
Fixing the freeze-out temperature of $T_f=150$~MeV reproduces the transverse momenta 
of identified particles at midrapidity.
To check the sensitivity of the results of the freeze-out temperature and viscosity, 
we have investigated also the cases when one of the parameter is modified from the default value
to $T_f=140$~MeV, $\eta/s=0.16$, or $\zeta/s=0.08$.
The calculations with a lower freeze-out temperature 
or with  an increased shear or bulk viscosity  give 
average transverse momenta within the range of the systematic errors quoted by the PHENIX Collaboration.
Admittedly, there is some model dependence on parameters, but it is weak, and the default parameters serve as an optimum choice. 

It has been noted that event-by-event hydrodynamics with lumpy initial conditions yields harder spectra
than hydrodynamics starting with averaged initial conditions \cite{Andrade:2008xh}. This effect follows 
from higher gradients in the lumpy initial condition. To compensate, i.e., to soften the spectra, one needs to run hydrodynamics  
for a shorter time, i.e., to higher freeze-out temperatures \cite{Bozek:2011ua}.

\section{Results \label{sec:res}}

The simulations presented in this section employ the experimental cuts in  the STAR \cite{Adams:2005ka}
($0.15~{\rm GeV} < p_T < 2~{\rm GeV}$) 
and PHENIX \cite{Adler:2003xq} ($0.2~{\rm GeV} < p_T < 2~{\rm GeV}$) 
analyses. In both cases $|\eta| <1$.
Our samples have 100 events at each considered centrality bin. 
These, involving the hydrodynamic evolution, are time-consuming to generate.
To increase the accuracy of the statistical hadronization, we generate 200  {\tt THERMINATOR} events for each hydro event. 

Our determination of centrality matches closely the experiment. In the case of STAR \cite{Adams:2005ka}, the multiplicity of 
generated charged particles in 
the window $|\eta| < 0.5$ is used to determine the centrality bins. In the case of PHENIX  \cite{Adler:2003xq}, where a combination of signals from the BBC 
and ZDC detectors is used, we simply take the number of wounded nucleons $N_w$ as the variable fixing the centrality.

\subsection{Fixed number of wounded nucleons \label{sec:fixed}}

\begin{figure}[tb]
\begin{center}
\includegraphics[angle=0,width=0.485 \textwidth]{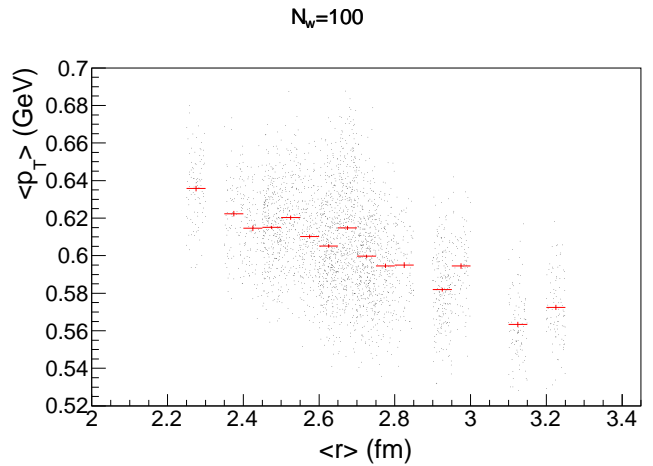} 
\end{center}
\vspace{-7mm}
\caption{(Color online)
Averaged transverse momentum as the function of the initial size $\langle r \rangle$
for events with a fixed number of wounded nucleons, $N_w=100$. Viscous ($3+1$)-D event-by-event hydrodynamics with $\eta/s=0.08$, 
$\zeta/s=0.04$, and $T_f=150$~MeV is used. The scattered small dots show 
$\langle p_T \rangle$ obtained in individual events, while the bars show the event-by-event averages $\langle \langle p_T \rangle \rangle$ in 
the selected bins of $\langle r \rangle$. The anticorrelation is apparent, with 
lower size $\langle r \rangle$ resulting in higher $\langle \langle p_T \rangle \rangle$.
\label{fig:anat}} 
\end{figure}

For better understanding, we begin the analysis for the event-by-event fluctuations by selecting a very narrow centrality 
class, with $N_w=100$. We run {\tt GLISSANDO} to generate the initial 
conditions, carry out our event-by-event hydrodynamics, and, finally, run {\tt THERMINATOR} and compute $\langle p_T \rangle$ in 
each event. As argued before \cite{Broniowski:2009fm}, the fluctuations of the initial condition manifest themselves 
in the fluctuations of the initial size $\langle r \rangle$. In Fig.~\ref{fig:anat} we plot the 
values of $\langle p_T \rangle$, histogrammed in bins of $\langle r \rangle$. Each point corresponds to one event, while the bars 
give the event-by-event average, $\langle \langle p_T \rangle \rangle$. We note a clear anticorrelation of  
$\langle \langle p_T \rangle \rangle$ and  $\langle  \langle 
r \rangle \rangle $. This shows that in a full-fledged 
event-by-event simulation the basis qualitative argument holds: for a squeezed initial the system expands with the larger flow 
velocity hand acquires a higher average transverse momentum, $\langle \langle p_T \rangle \rangle$, than for the stretched state.
The same effect can be observed when comparing case by case events generated from different initial conditions
(Fig. \ref{fig:photo}). The event with a squeezed initial density has a larger transverse flow and $\langle p_T \rangle$.

The fit to the histogram bars in Fig.~\ref{fig:anat} yields $\langle \langle p_T \rangle \rangle = 0.79 - 0.07 \langle \langle r \rangle \rangle$~GeV/fm, which 
in turn gives 
\begin{eqnarray}
\frac{d \langle \langle p_T \rangle \rangle}{d \langle \langle r \rangle \rangle} \simeq - 0.3 \frac{\langle \langle p_T \rangle \rangle}{\langle \langle r \rangle \rangle}
\end{eqnarray}
in the considered range. 
This result can be written as 
\begin{eqnarray}
\frac{\sigma(\langle p_T \rangle)}{\langle \langle p_T \rangle \rangle} \simeq 
0.3 \frac{\sigma(\langle r \rangle)}{\langle \langle r \rangle \rangle},
\end{eqnarray}
which may be compared to the estimate of Ref.~ \cite {Ollitrault:1991xx}, 
\begin{eqnarray}
\frac{\sigma(\langle p_T \rangle)}{\langle \langle p_T \rangle \rangle}
 = \frac{2\bar P}{\bar \epsilon} \frac{\sigma(\langle r \rangle)}{\langle \langle r \rangle \rangle}, \label{olli}
\end{eqnarray}
with $\bar P$ and $\bar \epsilon$ denoting the average pressure and energy density 
during the evolution of the system. Thus \mbox{${\bar P}/{\bar \epsilon} \sim 0.15$}, which is the right ball park 
for the applied equation of state \cite{Broniowski:2009te}. 
 
\subsection{Transverse momentum fluctuations vs. centrality \label{sec:rescen}}

\begin{figure}[tb]
\begin{center}
\includegraphics[angle=0,width=0.475 \textwidth]{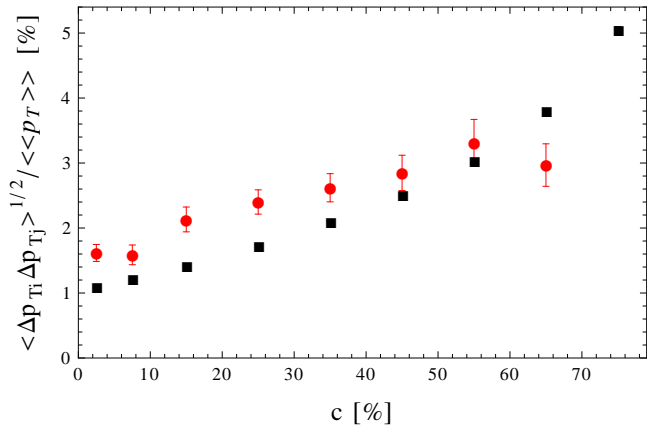} 
\end{center}
\vspace{-7mm}
\caption{(Color online) Comparison of the theoretical predictions for
 $\langle \Delta p_{Ti} \Delta p_{Tj} \rangle^{1/2}/\langle\langle p_T  \rangle\rangle$ 
(for $\sqrt{s_{NN}}=200$~GeV) 
to the experimental data extracted from the STAR Collaboration
\cite{Adams:2005ka} (squares). The dots correspond to simulation with event-by-event $(3+1)-$D viscous hydrodynamics with
our default parameters $T_f=150$~MeV, $\eta/s=0.08$, $\zeta/s=0.04$.
The statistical errors of the model simulation are obtained with the jackknife method. The experimental statistical errors are negligible. 
\label{fig:data}} 
\end{figure}  
 
\begin{figure}[tb]
\begin{center}
\includegraphics[angle=0,width=0.475 \textwidth]{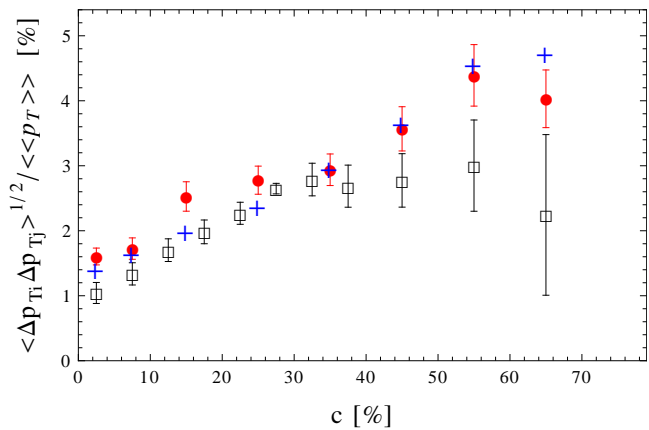} 
\end{center}
\vspace{-7mm}
\caption{(Color online) Comparison of the theoretical predictions for
 $\langle \Delta p_{Ti} \Delta p_{Tj} \rangle^{1/2}/\langle\langle p_T  \rangle\rangle$ 
(for $\sqrt{s_{NN}}=200$~GeV) 
to the experimental data from the PHENIX Collaboration
\cite{Adler:2003xq} (squares). The dots correspond to simulation with event-by-event $(3+1)$-D viscous hydrodynamics with 
our default parameters $T_f=150$~MeV, $\eta/s=0.08$, $\zeta/s=0.04$. 
The crosses indicate the approximate result from Ref.~\cite{Broniowski:2009fm} for perfect $(2+1)$-D hydrodynamics with averaged initial conditions from the mixed model. 
The statistical errors for the model simulations are obtained with the jackknife method. 
\label{fig:data_fen}} 
\end{figure}

Now we come to the main results of this paper. 
In order to compare to the data, we analyze the STAR correlation measure \cite{Adams:2005ka}, 
$\langle \Delta p_{Ti} \Delta p_{Tj} \rangle$, defined as 
\begin{eqnarray}
\langle \Delta p_{Ti} \Delta p_{Tj} \rangle \equiv \frac{1}{N_{\rm ev}} \sum_{k=1}^{N_{\rm ev}} \frac{C_k}{N_k(N_k-1)},
\end{eqnarray}
where $N_{\rm ev}$ is the number of events, $N_k$ the multiplicity in event $k$, and
\begin{eqnarray}
C_k=\sum_{i=1}^{N_k} \sum_{j=1,j\neq i}^{N_k} (p_i-\langle \langle p_T \rangle \rangle) 
(p_j-\langle \langle p_T \rangle \rangle), 
\label{star}
\end{eqnarray}
with \begin{eqnarray}
\langle \langle p_T \rangle \rangle = \frac{1}{N_{\rm ev}} \sum_{k=1}^{N_{\rm ev}} \langle p_T \rangle_k.
\end{eqnarray}
Introducing the mean momentum in event $k$, denoted by $\langle p_T \rangle_k$, we can transform 
\begin{eqnarray}
C_k= N_k (N_k-1) (\langle p_T \rangle_k - \langle \langle p_T \rangle \rangle )^2 - 
\sum_{i=1}^{N_k} (p_i- \langle p_T \rangle_k)^2, \nonumber \\
\end{eqnarray}
and rewrite
\begin{eqnarray}
\langle \Delta p_{Ti} \Delta p_{Tj} \rangle = 
\frac{N_{\rm ev}-1}{N_{\rm ev}} {\rm var}(\langle p_T \rangle) - 
\frac{1}{N_{\rm ev}} \sum_{k=1}^{N_{\rm ev}} \left [ \frac{{\rm var}_k ( p )}{N_k} \right ]. 
\nonumber \\ \label{my}
\end{eqnarray}
Thus the STAR correlation measure is the difference of two terms: one involving the variance of the mean momenta in events, and the
other being the event-averaged variance of the momentum in each event decided by the multiplicity of this event. Note that 
expression (\ref{my}) involves only single sums in a given event. As a matter of fact, the STAR analysis \cite{Adams:2005ka} replaces 
$\langle \langle p_T \rangle \rangle$ with the quantity $\langle p_T \rangle (N_{\rm charged})$, the average momentum as a function of the number of charged 
particles in the pseudorapidity bin $|\eta|<0.5$ -- the same as used to determine centrality. The function is obtained by a numerical 
fit to the results prior to the analysis of the correlations. The method slightly reduces the value of $\langle \Delta p_{Ti} \Delta p_{Tj} \rangle$.
We follow the same prescription.

Our results are shown in Fig.~\ref{fig:data}, where we compare the theoretical points (circles) to the experimental 
data from the STAR Collaboration \cite{Adams:2005ka} (squares).  At low centralities, the model calculations  
overshoot the data by about 50\%, yielding more $p_T$ fluctuations than needed. 
This conclusion supports the original findings of Ref.~\cite{Broniowski:2009fm} in the present state-of-the-art 
event-by-event treatment.  
We have checked for a few centrality bins that modifying the shear or bulk viscosity coefficients, the 
freeze-out temperature or the width of the smearing Gaussian in the initial conditions does not 
change the results at the level of the statistical errors of our calculations (see sect. \ref{sub:visc}).

Nevertheless, we note a proper magnitude of the effect and the 
correct dependence on centrality, Also, since the 
results of Fig.~\ref{fig:b} very weakly depend on $\sigma_{NN}$ \cite{Broniowski:2009fm}, 
with the expectation that the hydrodynamic ``push'' is similar at different collision energies, our results should 
weakly depend on the incident energy. This is a desired feature, as the STAR data \cite{Adams:2005ka} are very similar from 
$\sqrt{s_{NN}}=20$~GeV to 200~GeV. 

The statistical errors of the model simulations in Fig.~\ref{fig:data} are 
estimated with the jackknife method. Essentially, the relative error is 
equal to $1/\sqrt{2n}$, where $n=100$ is the number of the hydrodynamic events in the considered centrality class.

The PHENIX Collaboration \cite{Adler:2003xq} published results on the ratio of the $p_T$ fluctuations using the measure 
\begin{equation}
F_{p_T}=\frac{\omega_{p_T}^{\rm data}-\omega_{p_T}^{\rm mixed}}{\omega_{p_T}^{\rm mixed}} , \label{eq:fpt}
\end{equation}
where
\begin{equation}
\omega_{p_T}^{\rm data}=\frac{{\rm var}(\langle p_T \rangle )^{1/2}}{\langle\langle p_T \rangle\rangle}
\end{equation}
and $\omega_{p_T}^{mixed}$ is the same quantity obtained with mixed events.
For small dynamical fluctuations and sharp distributions in the 
multiplicity variable one can estimate\footnote{The relations between various popular 
correlations measures in this limit are discussed in the Appendix of Ref.~\cite{Broniowski:2006zz}. One of us (WB) thanks 
Jeff T.~Mitchell for the discussion concerning Eq.~(\ref{eq:fpt})}
\begin{equation}
\langle \Delta p_{Ti} \Delta p_{Tj}  \rangle \simeq 2 F_{p_T}\frac{{\rm var}(p_T)}{\langle N \rangle}, \label{eq:phm}
\end{equation}
where ${\rm var(p_T)}$ denotes the inclusive variance of the transverse momentum distribution, and $\langle N \rangle$ 
is the average multiplicity of the detected particles in the considered centrality class. The values of the quantities on the right-hand side 
of Eq.~(\ref{eq:phm}) are available from the PHENIX Collaboration web page associated with Ref.~\cite{Adler:2003xq}. We stress that 
the result Eq.~(\ref{eq:phm}) is approximate, but sufficiently accurate \cite{Broniowski:2006zz} for our purpose. A more direct comparison to 
the PHENIX data could be achieved with the mixing technique, however, this is beyond our reach due to a very limited number of the model events. 

The result of the analysis is shown in Fig.~\ref{fig:data_fen}, with similar conclusions as from Fig.~\ref{fig:data}, i.e., the model 
points are above the experiment.
We also show that the results of applying the event-by-event viscous hydrodynamics (dots) 
are very close to the approximate calculation of Ref.~\cite{Broniowski:2009fm} for perfect $(2+1)$-D hydrodynamics with 
averaged initial conditions (crosses).

\begin{figure}[tb]
\begin{center}
\includegraphics[angle=0,width=0.475 \textwidth]{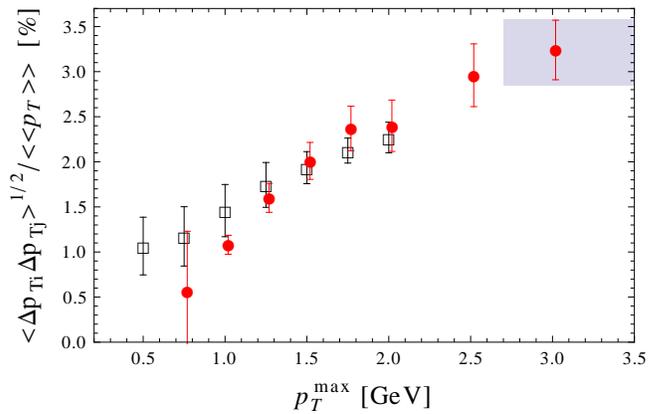} 
\end{center}
\vspace{-7mm}
\caption{(Color online) Dependence of 
 $\langle \Delta p_{Ti} \Delta p_{Tj} \rangle^{1/2}/\langle\langle p_T  \rangle\rangle$ 
(for $\sqrt{s_{NN}}=200$~GeV) on the upper transverse-momentum cut-off, compared
to the experimental data from the PHENIX Collaboration \cite{Adler:2003xq} for centrality $20-25$\% (squares). 
The dots correspond to simulation with event-by-event $(3+1)$-D viscous hydrodynamics with our default parameters $T_f=150$~MeV, $\eta/s=0.08$, $\zeta/s=0.04$. 
The statistical errors of the model simulation are obtained with the jackknife method. 
The shaded band in the upper right corner represents the error band of the 
result of the simulation with an infinite upper momentum cut-off.  
\label{fig:datapt}} 
\end{figure}

The dependence of  the fluctuation measure 
${\langle \Delta p_{Ti} \Delta p_{Tj}  \rangle^{1/2}}/{\langle \langle p_T \rangle \rangle}$ 
on the upper transverse-momentum cut-off has been measured by the PHENIX Collaboration~\cite{Adler:2003xq}. 
As can be seen in Fig. \ref{fig:datapt}, the fluctuations in the model increase with the cut-off, 
following closely the trend observed in the data. 
This cross-checks that the observed $p_T$-dependence of the transverse momentum fluctuations can be interpreted 
as a hydrodynamic flow effect.

\subsection{Dependence on model parameters \label{sub:visc}} 

In this section we investigate the dependence of our predictions on the model parameters, such as the viscosity 
coefficients of the medium, the freeze-out temperature, $T_f$, or the smoothing parameter, $w$. For this purpose
we have run simulations with various values of these parameters at fixed $N_w=100$.

\begin{figure}[tb]
\begin{center}
\includegraphics[angle=0,width=0.5 \textwidth]{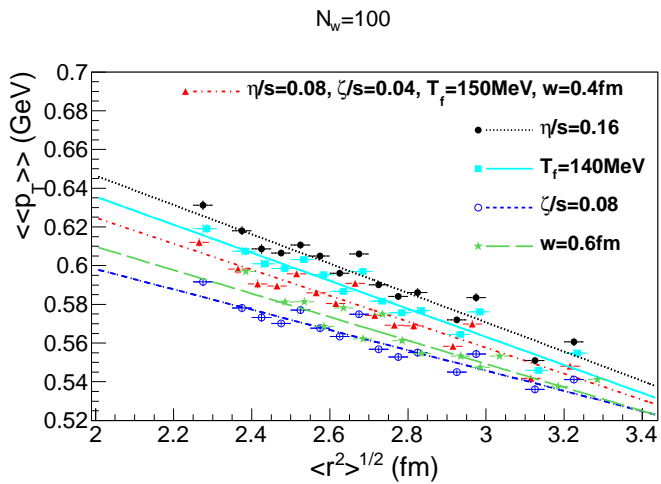} 
\end{center}
\vspace{-7mm}
\caption{(Color online) Dependence of $\langle \langle p_T \rangle \rangle$ on $\langle r \rangle$ for the variants
of the hydrodynamic evolution:  calculation with $\eta/s=0.08$, $\zeta/s=0.04$, $T_f=150$~MeV, and $w=0.4$~fm (default)
(triangles and dashed-dotted line), with $\eta/s=0.16$ (filled circles and dotted line), with $T_f=140$~MeV
(squares and solid line),  with $\zeta/s=0.08$ (open circles and dashed line), and with the width of the smearing $w=0.6$~fm (stars and long dashed line). 
The points with the errors bars 
represent the histogram of the average momentum $\langle \langle p_T \rangle \rangle$ as function of the r.m.s radius of the 
initial density of the event. The lines represent linear fits to the points. \label{fig:compare}} 
\end{figure}

In Fig.~\ref{fig:compare} we compare the dependence of $\langle p_T \rangle$ 
on $\langle r \rangle$ for several variants of viscous hydrodynamics.
The points correspond to mean $\langle \langle p_T \rangle \rangle$ in a given $\langle r \rangle$ bin, and the curves are linear 
fits to these points. 
Our default result is for the set of parameters $\eta/s=0.08$, $\zeta/s=0.04$, $T_f=150$~MeV, and $w=0.4$~fm,
indicated with the up-triangles and dot-dashed line in the plot. We then do our comparison
by changing one of the parameters:
the shear or bulk viscosity coefficient, the freeze-out temperature, or the smoothing parameter. 
As expected, increasing the shear viscosity or decreasing the freeze-out temperature leads to a 
hardening of the spectra, i.e, higher $\langle \langle p_T \rangle \rangle$. On the other hand, 
increasing bulk viscosity leads to a reduction of the effective pressure and a 
decrease of the average transverse momentum. Increasing the  smearing width of the initial density distribution 
yields smaller gradients, and reduces the transverse push. The described behavior holds bin-by-bin in the 
$\langle r \rangle$ variable giving the size of the initial geometry.

At the same time we note that the slope of the dependence of $\langle \langle p_T \rangle \rangle$ on  $\langle r \rangle$ 
changes as well. For all the studied cases it turns out that to a high accuracy
\begin{eqnarray}
\frac{d\langle p_T \rangle}{d\langle r \rangle} \frac{\langle r \rangle}{\langle p_T \rangle}  
 \simeq 0.31.
\end{eqnarray}   
As a result, according to the arguments of Sec.~\ref{sec:fixed}, the scaled measure 
$\langle \Delta p_{Ti} \Delta p_{Tj} \rangle/\langle \langle p_T\rangle \rangle^2$ is hardly modified.
Increasing the transverse pressure or the local gradients gives a larger transverse flow and, simultaneously, larger
fluctuations, such that the scaled fluctuations  of the transverse momentum practically 
do not depend on viscosity, the freeze-out temperature, or the smearing of the initial conditions.
We thus find that the scaled $p_T$-fluctuations are dominated by the fluctuations of the transverse size
of the initial fireball.
This feature makes the scaled measure particularly suitable for constraining the models of the initial phase.
In a similar way, the directed flow at central rapidity has been proposed as a tool to limit the dipole deformation 
of the fireball predicted by different models of the initial state \cite{Retinskaya:2012ky}. 

\subsection{Other effects}

In peripheral collisions the particle emission can take place in the thermalized, collectively expanding 
core, as well as in the outer corona, where rescattering is small 
\cite{Hohne:2006ks,Becattini:2008ya,Bozek:2005eu,Werner:2007bf}.
In the following we estimate the transverse momentum fluctuations in the case where the particles are emitted 
from these two sources, the core and the corona. Different definitions of the dense core are possible; we use the
prescription that a wounded nucleon belongs to the core if it collides more than once. This choice of the separation  between
the core and the corona describes the centrality dependence of the strangeness production and of 
the effective slopes of the particle spectra \cite{Becattini:2008ya,Bozek:2008zw}.
The particle density at central rapidity is a sum of the contributions from the corona and the core,
\begin{equation}
\frac{dN_{\rm charged}}{d\eta}=\frac{dN_{\rm charged}^{NN}}{d\eta}( N_{\rm corona}  +\beta N_{\rm core}) , 
\end{equation}
with $N_{\rm core}+N_{\rm corona}=N_w$.
To reproduce the centrality dependence observed experimentally we choose the parameter $\beta=1.75$, which
effectively describes the enhanced production in the thermalized  matter \cite{Bozek:2005eu}. The average transverse momentum,
\begin{eqnarray}
\langle \langle p_T \rangle\rangle= (1-c) \langle \langle p_T^{pp} \rangle\rangle + c 
\langle \langle p_T^{\rm core} \rangle\rangle,
\end{eqnarray}
is a combination of the average momentum in a $NN$-collision and of the 
average transverse momentum of
particles emitted from the core. The fraction of particles emitted from the core is $c=\beta N_{\rm core}/(\beta N_{\rm core}+N _{\rm corona})$.
By neglecting the fluctuations of the number of nucleons in the core, we get 
\begin{eqnarray}
\langle \Delta p_{Ti} \Delta p_{Tj}  \rangle & \simeq& \frac{1}{\langle N_{\rm corona}\rangle} (1-c)^2 \langle \Delta p_{Ti} \Delta p_{Tj}   
\rangle^{NN} \nonumber \\ & +&  c^2(1-c)^2 ( \langle \langle p_T^{NN} \rangle\rangle -
\langle \langle p_T^{\rm core} \rangle\rangle)^2 \nonumber \\ 
&+ & c^2 \langle \Delta p_{Ti} \Delta p_{Tj}  \rangle \rangle^{\rm core} .
\label{eq:core}
\end{eqnarray}
The first term is a contribution from the $N_{\rm corona}$ independent $NN$ sources, the second term comes from
the difference of the transverse momenta from the two sources, and the third term is a contribution from the 
hydrodynamically expanding core. 
We use the PHENIX data \cite{Adler:2003xq} on the average transverse momentum and its fluctuations in the $pp$-collisions, and take for 
$\langle \Delta p_{Ti} \Delta p_{Tj} \rangle^{\rm core}$ the results from the hydrodynamic model calculation. 
With all elements of Eq.~(\ref{eq:core}) combined,
we find that 
$\langle \Delta p_{Ti} \Delta p_{Tj} \rangle^{1/2}/\langle \langle p_T \rangle \rangle $ is changed very little 
compared to the results of Sec.~\ref{sec:rescen}.
For the most central collisions, where the corona contribution is tiny, naturally the effect is negligible. 
In peripheral collisions ($c=60-70\%$) all terms of Eq.~(\ref{eq:core}) contribute to the transverse momentum fluctuations. However, 
in the last term the reduction 
of the core due to the $(1-c)^2$ factor is compensated with increased fluctuations $\sigma(\langle r \rangle)/\langle r \rangle$, such that 
with all terms combined the change is at the level of 10\%. At intermediate centralities the reduction effect is at a similar level. 
Therefore the core-corona model does not improve nor deteriorate the agreement with the experimental data.

We have also checked that imposing a finite detector acceptance, by simply accepting a simulated particle with the typical probability of 
50\%, does not alter the results for $\langle \Delta p_{Ti} \Delta p_{Tj}\rangle /\langle \langle p_T \rangle \rangle$. This is a feature of the 
scaled $p_T$ fluctuation measure \cite{Adler:2003xq,Adamova:2008sx}.

Finally, we have estimated the possible effect of the global transverse-momentum conservation, not 
implemented in the standard simulations of the statistical hadronization with {\tt THERMINATOR}. This can be 
approximately achieved by accepting only those events which have limited total transverse momentum. 
Specifically, we consider the quantity $P^2=(\sum p_{i,x})^2 + (\sum p_{i,y})^2$ in a given event and include the event 
for the further analysis when
$P$ is less than a specified value, which is gradually decreased. With our
statistics we are able to reduce the limit for $P$ down to $15$~GeV 
(there are a few hundred of particles in the event), which leaves about 5\% from all (unconstrained) events for the most central 
case. No noticeable effect is detected, therefore 
the considered correlation measure is not sensitive to the global transverse-momentum conservation.

\section{Conclusions \label{sec:concl}}

The initial shape and the volume of the fireball fluctuate due to the random nature of the Glauber approach. 
As is well known, the subsequent hydrodynamic evolution carries over the asymmetry of the shape of
the fireball into anisotropies of the particle spectra. A similar mechanism, analyzed in detail in this work,
transmits the event-by-event fluctuations of the transverse size of the fireball into the fluctuations of the average 
transverse momentum in each event, as identified in \cite{Broniowski:2009fm}.
Here are the main findings of our analysis:

\begin{enumerate}
 
\item The state-of-the-art event-by-event viscous $(3+1)$-D hydrodynamic calculations with fluctuating initial conditions confirm that 
fluctuations of the mean transverse momentum in each event are generated from the fluctuations of the initial geometry. 

\item The amount of scaled transverse momentum fluctuations is determined by the scaled fluctuations of the transverse size of the fireball.
We observe an anticorrelation of the initial size of the fireball and of the transverse momentum 
generated in an event. The expansion of a source of larger extent yields smaller $p_T$ then in the case of a squeezed source, and vice versa.

\item Hydrodynamic expansion is applied to an 
ensemble of events, corresponding to centralities $0-70\%$ in  Au-Au collisions at $\sqrt{s_{NN}}=200$~GeV.
We find a similar magnitude and centrality dependence of the scaled momentum fluctuations 
$\langle \Delta p_{Ti} \Delta p_{Tj} \rangle /\langle \langle p_T \rangle \rangle$ 
as in the STAR \cite{Adams:2005ka} and PHENIX experiments \cite{Adler:2003xq}. 

\item The dependence of the results on the upper cut-off for the transverse momentum of the particles agrees nicely with the data 
from the PHENIX Collaboration \cite{Adler:2003xq}.

\item However, the initial density from the mixed model (wounded nucleons with an admixture of binary collisions), tuned to 
reproduce the particle multiplicities, yields a visible overprediction of 
the observed value of $\langle \Delta p_{Ti} \Delta p_{Tj} \rangle/ \langle \langle p_T \rangle \rangle$ in the whole centrality range. For most central 
events the overprediction is at the level of $50\%$, while it gets relatively closer to the data with increased centrality. 

\item Hydrodynamic expansion yields a stronger transverse push and, simultaneously, stronger $p_T$-fluctuations 
when the shear viscosity is increased, the freeze-out temperature is lowered, or if the 
bulk viscosity is lowered. However, the scaled fluctuation measure $\langle \Delta p_{Ti} \Delta p_{Tj} \rangle /\langle \langle p_T \rangle \rangle$ 
shows very little changes with these modifications of the physical parameters. 

\item Predictions of our approach remain essentially unchanged when the core-corona mechanism of particle emission is incorporated. 
Other effects, such as the transverse-momentum conservation or the finite detector acceptance do not affect the results, either.

\end{enumerate}

~

The above points indicate that the identified ``geometric'' mechanism of generating 
the transverse-momentum fluctuations from the initial Glauber-like model is, on the one hand, very important, 
easily reproducing the size of the effect and
catching the basic features of the data, on the other hand, it is somewhat too strong.
That hints on an improvement of the popular Glauber approach of the initial phase. 
We recall that the calculations using the averaged initial conditions \cite{Broniowski:2009fm}
show that the fluctuations of the initial size are reduced if 
the density in the fireball is determined with the wounded nucleons only, i.e.,  without the 
admixture of binary collisions. At the same time, however, the model with the wounded nucleons only fails 
to generate the proper multiplicity dependence on centrality, thus is less realistic.  Moreover, hydrodynamic fluctuations in the evolution could add another source of
$\langle p_T\rangle$  fluctuations \cite{Florchinger:2011qf,Kapusta:2012gm}.
Thus, it remains a challenge to understand in detail the earliest phase of the collision and 
reproduce in a {\em uniform} way the rich collection of the one-body and the correlation data, 
including also the harmonic flow. 

\begin{acknowledgments}
Supported by Polish Ministry of Science and Higher Education, grant N~N202~263438, and National Science 
Centre, grant DEC-2011/01/D/ST2/00772.  Part of the numerical calculations were made on the Cracow Cloud 
One cluster.
\end{acknowledgments}

\bibliography{hydr}

\begin{thebibliography}{100}%
\makeatletter
\providecommand \@ifxundefined [1]{%
 \ifx #1\undefined \expandafter \@firstoftwo
 \else \expandafter \@secondoftwo
\fi
}%
\providecommand \@ifnum [1]{%
 \ifnum #1\expandafter \@firstoftwo
 \else \expandafter \@secondoftwo
\fi
}%
\providecommand \enquote [1]{``#1''}%
\providecommand \bibnamefont  [1]{#1}%
\providecommand \bibfnamefont [1]{#1}%
\providecommand \citenamefont [1]{#1}%
\providecommand\href[0]{\@sanitize\@href}%
\providecommand\@href[1]{\endgroup\@@startlink{#1}\endgroup\@@href}%
\providecommand\@@href[1]{#1\@@endlink}%
\providecommand \@sanitize [0]{\begingroup\catcode`\&12\catcode`\#12\relax}%
\@ifxundefined \pdfoutput {\@firstoftwo}{%
 \@ifnum{\z@=\pdfoutput}{\@firstoftwo}{\@secondoftwo}%
}{%
 \providecommand\@@startlink[1]{\leavevmode\special{html:<a href="#1">}}%
 \providecommand\@@endlink[0]{\special{html:</a>}}%
}{%
 \providecommand\@@startlink[1]{%
  \leavevmode
  \pdfstartlink
   attr{/Border[0 0 1 ]/H/I/C[0 1 1]}%
   user{/Subtype/Link/A<</Type/Action/S/URI/URI(#1)>>}%
  \relax
 }%
 \providecommand\@@endlink[0]{\pdfendlink}%
}%
\providecommand \url  [0]{\begingroup\@sanitize \@url }%
\providecommand \@url [1]{\endgroup\@href {#1}{\urlprefix}}%
\providecommand \urlprefix [0]{URL }%
\providecommand \Eprint[0]{\href }%
\@ifxundefined \urlstyle {%
  \providecommand \doi [1]{doi:\discretionary{}{}{}#1}%
}{%
  \providecommand \doi [0]{doi:\discretionary{}{}{}\begingroup
  \urlstyle{rm}\Url }%
}%
\providecommand \doibase [0]{http://dx.doi.org/}%
\providecommand \Doi[1]{\href{\doibase#1}}%
\providecommand \bibAnnote [3]{%
  \BibitemShut{#1}%
  \begin{quotation}\noindent
    \textsc{Key:}\ #2\\\textsc{Annotation:}\ #3%
  \end{quotation}%
}%
\providecommand \bibAnnoteFile [2]{%
  \IfFileExists{#2}{\bibAnnote {#1} {#2} {\input{#2}}}{}%
}%
\providecommand \typeout [0]{\immediate \write \m@ne }%
\providecommand \selectlanguage [0]{\@gobble}%
\providecommand \bibinfo [0]{\@secondoftwo}%
\providecommand \bibfield [0]{\@secondoftwo}%
\providecommand \translation [1]{[#1]}%
\providecommand \BibitemOpen[0]{}%
\providecommand \bibitemStop [0]{}%
\providecommand \bibitemNoStop [0]{.\EOS\space}%
\providecommand \EOS [0]{\spacefactor3000\relax}%
\providecommand \BibitemShut [1]{\csname bibitem#1\endcsname}%
\bibitem{Broniowski:2009fm}%
  \BibitemOpen
  \bibfield{author}{%
  \bibinfo {author} {\bibfnamefont{W.}~\bibnamefont{Broniowski}}, \bibinfo
  {author} {\bibfnamefont{M.}~\bibnamefont{Chojnacki}},\ and\ \bibinfo {author}
  {\bibfnamefont{L.}~\bibnamefont{Obara}},\ }%
  \bibfield{journal}{%
  \Doi{10.1103/PhysRevC.80.051902}{\bibinfo {journal} {Phys. Rev.}}\ }%
  \textbf{\bibinfo {volume} {C80}},\ \bibinfo {pages} {051902} (\bibinfo {year}
  {2009})%
  \bibAnnoteFile{NoStop}{Broniowski:2009fm}%
\bibitem{Broniowski:2007nz}%
  \BibitemOpen
  \bibfield{author}{%
  \bibinfo {author} {\bibfnamefont{W.}~\bibnamefont{Broniowski}}, \bibinfo
  {author} {\bibfnamefont{M.}~\bibnamefont{Rybczy\'nski}},\ and\ \bibinfo
  {author} {\bibfnamefont{P.}~\bibnamefont{Bo\.zek}},\ }%
  \bibfield{journal}{%
  \Doi{10.1016/j.cpc.2008.07.016}{\bibinfo {journal} {Comput. Phys. Commun.}}\
  }%
  \textbf{\bibinfo {volume} {180}},\ \bibinfo {pages} {69} (\bibinfo {year}
  {2009})%
  \bibAnnoteFile{NoStop}{Broniowski:2007nz}%
\bibitem{Bozek:2011ua}%
  \BibitemOpen
  \bibfield{author}{%
  \bibinfo {author} {\bibfnamefont{P.}~\bibnamefont{Bo\.zek}},\ }%
  \bibfield{journal}{%
  \Doi{10.1103/PhysRevC.85.034901}{\bibinfo {journal} {Phys. Rev.}}\ }%
  \textbf{\bibinfo {volume} {C85}},\ \bibinfo {pages} {034901} (\bibinfo {year}
  {2012})%
  \bibAnnoteFile{NoStop}{Bozek:2011ua}%
\bibitem{Bozek:2011if}%
  \BibitemOpen
  \bibfield{author}{%
  \bibinfo {author} {\bibfnamefont{P.}~\bibnamefont{Bo\.zek}},\ }%
  \bibfield{journal}{%
  \bibinfo {journal} {Phys. Rev.}\ }%
  \textbf{\bibinfo {volume} {C85}},\ \bibinfo {pages} {014911} (\bibinfo {year}
  {2012})%
  \bibAnnoteFile{NoStop}{Bozek:2011if}%
\bibitem{Kisiel:2005hn}%
  \BibitemOpen
  \bibfield{author}{%
  \bibinfo {author} {\bibfnamefont{A.}~\bibnamefont{Kisiel}}, \bibinfo {author}
  {\bibfnamefont{T.}~\bibnamefont{{Ta\l{}u\'c}}}, \bibinfo {author}
  {\bibfnamefont{W.}~\bibnamefont{Broniowski}},\ and\ \bibinfo {author}
  {\bibfnamefont{W.}~\bibnamefont{Florkowski}},\ }%
  \bibfield{journal}{%
  \bibinfo {journal} {Comput. Phys. Commun.}\ }%
  \textbf{\bibinfo {volume} {174}},\ \bibinfo {pages} {669} (\bibinfo {year}
  {2006})%
  \bibAnnoteFile{NoStop}{Kisiel:2005hn}%
\bibitem{Chojnacki:2011hb}%
  \BibitemOpen
  \bibfield{author}{%
  \bibinfo {author} {\bibfnamefont{M.}~\bibnamefont{Chojnacki}}, \bibinfo
  {author} {\bibfnamefont{A.}~\bibnamefont{Kisiel}}, \bibinfo {author}
  {\bibfnamefont{W.}~\bibnamefont{Florkowski}},\ and\ \bibinfo {author}
  {\bibfnamefont{W.}~\bibnamefont{Broniowski}},\ }%
  \bibfield{journal}{%
  \Doi{10.1016/j.cpc.2011.11.018}{\bibinfo {journal} {Comput.Phys.Commun.}}\ }%
  \textbf{\bibinfo {volume} {183}},\ \bibinfo {pages} {746} (\bibinfo {year}
  {2012})%
  \bibAnnoteFile{NoStop}{Chojnacki:2011hb}%
\bibitem{Gazdzicki:1992ri}%
  \BibitemOpen
  \bibfield{author}{%
  \bibinfo {author} {\bibfnamefont{M.}~\bibnamefont{Ga\'zdzicki}}\ and\
  \bibinfo {author} {\bibfnamefont{S.}~\bibnamefont{Mr\'owczy\'nski}},\ }%
  \bibfield{journal}{%
  \bibinfo {journal} {Z. Phys.}\ }%
  \textbf{\bibinfo {volume} {C54}},\ \bibinfo {pages} {127} (\bibinfo {year}
  {1992})%
  \bibAnnoteFile{NoStop}{Gazdzicki:1992ri}%
\bibitem{Stodolsky:1995ds}%
  \BibitemOpen
  \bibfield{author}{%
  \bibinfo {author} {\bibfnamefont{L.}~\bibnamefont{Stodolsky}},\ }%
  \bibfield{journal}{%
  \bibinfo {journal} {Phys. Rev. Lett.}\ }%
  \textbf{\bibinfo {volume} {75}},\ \bibinfo {pages} {1044} (\bibinfo {year}
  {1995})%
  \bibAnnoteFile{NoStop}{Stodolsky:1995ds}%
\bibitem{Shuryak:1997yj}%
  \BibitemOpen
  \bibfield{author}{%
  \bibinfo {author} {\bibfnamefont{E.~V.}\ \bibnamefont{Shuryak}},\ }%
  \bibfield{journal}{%
  \bibinfo {journal} {Phys. Lett.}\ }%
  \textbf{\bibinfo {volume} {B423}},\ \bibinfo {pages} {9} (\bibinfo {year}
  {1998})%
  \bibAnnoteFile{NoStop}{Shuryak:1997yj}%
\bibitem{Mrowczynski:1997kz}%
  \BibitemOpen
  \bibfield{author}{%
  \bibinfo {author} {\bibfnamefont{S.}~\bibnamefont{Mr\'owczy\'nski}},\ }%
  \bibfield{journal}{%
  \bibinfo {journal} {Phys. Lett.}\ }%
  \textbf{\bibinfo {volume} {B430}},\ \bibinfo {pages} {9} (\bibinfo {year}
  {1998})%
  \bibAnnoteFile{NoStop}{Mrowczynski:1997kz}%
\bibitem{Liu:1998xf}%
  \BibitemOpen
  \bibfield{author}{%
  \bibinfo {author} {\bibfnamefont{F.}~\bibnamefont{Liu}}, \bibinfo {author}
  {\bibfnamefont{A.}~\bibnamefont{Tai}}, \bibinfo {author}
  {\bibfnamefont{M.}~\bibnamefont{Gazdzicki}},\ and\ \bibinfo {author}
  {\bibfnamefont{R.}~\bibnamefont{Stock}},\ }%
  \bibfield{journal}{%
  \Doi{10.1007/s100529900002}{\bibinfo {journal} {Eur.Phys.J.}}\ }%
  \textbf{\bibinfo {volume} {C8}},\ \bibinfo {pages} {649} (\bibinfo {year}
  {1999})%
  \bibAnnoteFile{NoStop}{Liu:1998xf}%
\bibitem{Voloshin:1999yf}%
  \BibitemOpen
  \bibfield{author}{%
  \bibinfo {author} {\bibfnamefont{S.~A.}\ \bibnamefont{Voloshin}}, \bibinfo
  {author} {\bibfnamefont{V.}~\bibnamefont{Koch}},\ and\ \bibinfo {author}
  {\bibfnamefont{H.~G.}\ \bibnamefont{Ritter}},\ }%
  \bibfield{journal}{%
  \bibinfo {journal} {Phys. Rev.}\ }%
  \textbf{\bibinfo {volume} {C60}},\ \bibinfo {pages} {024901} (\bibinfo {year}
  {1999})%
  \bibAnnoteFile{NoStop}{Voloshin:1999yf}%
\bibitem{Baym:1999up}%
  \BibitemOpen
  \bibfield{author}{%
  \bibinfo {author} {\bibfnamefont{G.}~\bibnamefont{Baym}}\ and\ \bibinfo
  {author} {\bibfnamefont{H.}~\bibnamefont{Heiselberg}},\ }%
  \bibfield{journal}{%
  \bibinfo {journal} {Phys. Lett.}\ }%
  \textbf{\bibinfo {volume} {B469}},\ \bibinfo {pages} {7} (\bibinfo {year}
  {1999})%
  \bibAnnoteFile{NoStop}{Baym:1999up}%
\bibitem{Voloshin:2001ei}%
  \BibitemOpen
  \bibfield{author}{%
  \bibinfo {author} {\bibfnamefont{S.~A.}\ \bibnamefont{Voloshin}} (\bibinfo
  {collaboration} {STAR}),\ }%
  \bibfield{journal}{%
  \Doi{10.1063/1.1469997}{\bibinfo {journal} {AIP Conf. Proc.}}\ }%
  \textbf{\bibinfo {volume} {610}},\ \bibinfo {pages} {591} (\bibinfo {year}
  {2001})%
  \bibAnnoteFile{NoStop}{Voloshin:2001ei}%
\bibitem{Korus:2001au}%
  \BibitemOpen
  \bibfield{author}{%
  \bibinfo {author} {\bibfnamefont{R.}~\bibnamefont{Korus}}, \bibinfo {author}
  {\bibfnamefont{S.}~\bibnamefont{Mr\'owczy\'nski}}, \bibinfo {author}
  {\bibfnamefont{M.}~\bibnamefont{Rybczy\'nski}},\ and\ \bibinfo {author}
  {\bibfnamefont{Z.}~\bibnamefont{W\l{}odarczyk}},\ }%
  \bibfield{journal}{%
  \bibinfo {journal} {Phys. Rev.}\ }%
  \textbf{\bibinfo {volume} {C64}},\ \bibinfo {pages} {054908} (\bibinfo {year}
  {2001})%
  \bibAnnoteFile{NoStop}{Korus:2001au}%
\bibitem{Gavin:2003cb}%
  \BibitemOpen
  \bibfield{author}{%
  \bibinfo {author} {\bibfnamefont{S.}~\bibnamefont{Gavin}},\ }%
  \bibfield{journal}{%
  \bibinfo {journal} {Phys. Rev. Lett.}\ }%
  \textbf{\bibinfo {volume} {92}},\ \bibinfo {pages} {162301} (\bibinfo {year}
  {2004})%
  \bibAnnoteFile{NoStop}{Gavin:2003cb}%
\bibitem{DiasdeDeus:2003ei}%
  \BibitemOpen
  \bibfield{author}{%
  \bibinfo {author} {\bibfnamefont{J.}~\bibnamefont{Dias~de Deus}}, \bibinfo
  {author} {\bibfnamefont{E.}~\bibnamefont{Ferreiro}}, \bibinfo {author}
  {\bibfnamefont{C.}~\bibnamefont{Pajares}},\ and\ \bibinfo {author}
  {\bibfnamefont{R.}~\bibnamefont{Ugoccioni}},\ }%
  \bibfield{journal}{%
  \Doi{10.1140/epjc/s2005-02127-y}{\bibinfo {journal} {Eur.Phys.J.}}\ }%
  \textbf{\bibinfo {volume} {C40}},\ \bibinfo {pages} {229} (\bibinfo {year}
  {2005})%
  \bibAnnoteFile{NoStop}{DiasdeDeus:2003ei}%
\bibitem{Voloshin:2004th}%
  \BibitemOpen
  \bibfield{author}{%
  \bibinfo {author} {\bibfnamefont{S.~A.}\ \bibnamefont{Voloshin}},\ }%
  \bibfield{journal}{%
  \Doi{10.1016/j.nuclphysa.2004.12.053}{\bibinfo {journal} {Nucl.Phys.}}\ }%
  \textbf{\bibinfo {volume} {A749}},\ \bibinfo {pages} {287} (\bibinfo {year}
  {2005})%
  \bibAnnoteFile{NoStop}{Voloshin:2004th}%
\bibitem{Mrowczynski:2004cg}%
  \BibitemOpen
  \bibfield{author}{%
  \bibinfo {author} {\bibfnamefont{S.}~\bibnamefont{Mr\'owczy\'nski}}, \bibinfo
  {author} {\bibfnamefont{M.}~\bibnamefont{Rybczy\'nski}},\ and\ \bibinfo
  {author} {\bibfnamefont{Z.}~\bibnamefont{W\l{}odarczyk}},\ }%
  \bibfield{journal}{%
  \Doi{10.1103/PhysRevC.70.054906}{\bibinfo {journal} {Phys.Rev.}}\ }%
  \textbf{\bibinfo {volume} {C70}},\ \bibinfo {pages} {054906} (\bibinfo {year}
  {2004})%
  \bibAnnoteFile{NoStop}{Mrowczynski:2004cg}%
\bibitem{AbdelAziz:2005wc}%
  \BibitemOpen
  \bibfield{author}{%
  \bibinfo {author} {\bibfnamefont{M.}~\bibnamefont{Abdel-Aziz}}\ and\ \bibinfo
  {author} {\bibfnamefont{S.}~\bibnamefont{Gavin}},\ }%
  \bibfield{journal}{%
  \Doi{10.1016/j.nuclphysa.2006.06.100}{\bibinfo {journal} {Nucl.Phys.}}\ }%
  \textbf{\bibinfo {volume} {A774}},\ \bibinfo {pages} {623} (\bibinfo {year}
  {2006})%
  \bibAnnoteFile{NoStop}{AbdelAziz:2005wc}%
\bibitem{Broniowski:2005ae}%
  \BibitemOpen
  \bibfield{author}{%
  \bibinfo {author} {\bibfnamefont{W.}~\bibnamefont{Broniowski}}, \bibinfo
  {author} {\bibfnamefont{B.}~\bibnamefont{Hiller}}, \bibinfo {author}
  {\bibfnamefont{W.}~\bibnamefont{Florkowski}},\ and\ \bibinfo {author}
  {\bibfnamefont{P.}~\bibnamefont{Bo\.zek}},\ }%
  \bibfield{journal}{%
  \Doi{10.1016/j.physletb.2006.02.056}{\bibinfo {journal} {Phys.Lett.}}\ }%
  \textbf{\bibinfo {volume} {B635}},\ \bibinfo {pages} {290} (\bibinfo {year}
  {2006})%
  \bibAnnoteFile{NoStop}{Broniowski:2005ae}%
\bibitem{Prindle:2006zz}%
  \BibitemOpen
  \bibfield{author}{%
  \bibinfo {author} {\bibfnamefont{D.~J.}\ \bibnamefont{Prindle}}\ and\
  \bibinfo {author} {\bibfnamefont{T.~A.}\ \bibnamefont{Trainor}} (\bibinfo
  {collaboration} {STAR Collaboration}),\ }%
  \bibfield{journal}{%
  \bibinfo {journal} {PoS}\ }%
  \textbf{\bibinfo {volume} {CFRNC2006}},\ \bibinfo {pages} {007} (\bibinfo
  {year} {2006})%
  \bibAnnoteFile{NoStop}{Prindle:2006zz}%
\bibitem{Gavin:2006xd}%
  \BibitemOpen
  \bibfield{author}{%
  \bibinfo {author} {\bibfnamefont{S.}~\bibnamefont{Gavin}}\ and\ \bibinfo
  {author} {\bibfnamefont{M.}~\bibnamefont{Abdel-Aziz}},\ }%
  \bibfield{journal}{%
  \Doi{10.1103/PhysRevLett.97.162302}{\bibinfo {journal} {Phys.Rev.Lett.}}\ }%
  \textbf{\bibinfo {volume} {97}},\ \bibinfo {pages} {162302} (\bibinfo {year}
  {2006})%
  \bibAnnoteFile{NoStop}{Gavin:2006xd}%
\bibitem{Sharma:2008qr}%
  \BibitemOpen
  \bibfield{author}{%
  \bibinfo {author} {\bibfnamefont{M.}~\bibnamefont{Sharma}}\ and\ \bibinfo
  {author} {\bibfnamefont{C.~A.}\ \bibnamefont{Pruneau}},\ }%
  \bibfield{journal}{%
  \Doi{10.1103/PhysRevC.79.024905}{\bibinfo {journal} {Phys.Rev.}}\ }%
  \textbf{\bibinfo {volume} {C79}},\ \bibinfo {pages} {024905} (\bibinfo {year}
  {2009})%
  \bibAnnoteFile{NoStop}{Sharma:2008qr}%
\bibitem{Mrowczynski:2009wk}%
  \BibitemOpen
  \bibfield{author}{%
  \bibinfo {author} {\bibfnamefont{S.}~\bibnamefont{Mr\'owczy\'nski}},\ }%
  \bibfield{journal}{%
  \bibinfo {journal} {Acta Phys.Polon.}\ }%
  \textbf{\bibinfo {volume} {B40}},\ \bibinfo {pages} {1053} (\bibinfo {year}
  {2009})%
  \bibAnnoteFile{NoStop}{Mrowczynski:2009wk}%
\bibitem{Hama:2009pk}%
  \BibitemOpen
  \bibfield{author}{%
  \bibinfo {author} {\bibfnamefont{Y.}~\bibnamefont{Hama}}, \bibinfo {author}
  {\bibfnamefont{R.~P.~G.}\ \bibnamefont{Andrade}}, \bibinfo {author}
  {\bibfnamefont{F.}~\bibnamefont{Grassi}}, \bibinfo {author}
  {\bibfnamefont{W.~L.}\ \bibnamefont{Qian}},\ and\ \bibinfo {author}
  {\bibfnamefont{T.}~\bibnamefont{Kodama}},\ }%
  \bibfield{journal}{%
  \bibinfo {journal} {Acta Phys. Polon.}\ }%
  \textbf{\bibinfo {volume} {B40}},\ \bibinfo {pages} {931} (\bibinfo {year}
  {2009})%
  \bibAnnoteFile{NoStop}{Hama:2009pk}%
\bibitem{Adams:2003uw}%
  \BibitemOpen
  \bibfield{author}{%
  \bibinfo {author} {\bibfnamefont{J.}~\bibnamefont{Adams}} \emph{et~al.}
  (\bibinfo {collaboration} {STAR Collaboration}),\ }%
  \bibfield{journal}{%
  \Doi{10.1103/PhysRevC.71.064906}{\bibinfo {journal} {Phys.Rev.}}\ }%
  \textbf{\bibinfo {volume} {C71}},\ \bibinfo {pages} {064906} (\bibinfo {year}
  {2005})%
  \bibAnnoteFile{NoStop}{Adams:2003uw}%
\bibitem{Adamova:2003pz}%
  \BibitemOpen
  \bibfield{author}{%
  \bibinfo {author} {\bibfnamefont{D.}~\bibnamefont{Adamova}} \emph{et~al.}
  (\bibinfo {collaboration} {CERES Collaboration}),\ }%
  \bibfield{journal}{%
  \Doi{10.1016/j.nuclphysa.2003.07.018}{\bibinfo {journal} {Nucl.Phys.}}\ }%
  \textbf{\bibinfo {volume} {A727}},\ \bibinfo {pages} {97} (\bibinfo {year}
  {2003})%
  \bibAnnoteFile{NoStop}{Adamova:2003pz}%
\bibitem{Adler:2003xq}%
  \BibitemOpen
  \bibfield{author}{%
  \bibinfo {author} {\bibfnamefont{S.}~\bibnamefont{Adler}} \emph{et~al.}
  (\bibinfo {collaboration} {PHENIX Collaboration}),\ }%
  \bibfield{journal}{%
  \Doi{10.1103/PhysRevLett.93.092301}{\bibinfo {journal} {Phys.Rev.Lett.}}\ }%
  \textbf{\bibinfo {volume} {93}},\ \bibinfo {pages} {092301} (\bibinfo {year}
  {2004})%
  \bibAnnoteFile{NoStop}{Adler:2003xq}%
\bibitem{Anticic:2003fd}%
  \BibitemOpen
  \bibfield{author}{%
  \bibinfo {author} {\bibfnamefont{T.}~\bibnamefont{Anticic}} \emph{et~al.}
  (\bibinfo {collaboration} {NA49 Collaboration}),\ }%
  \bibfield{journal}{%
  \Doi{10.1103/PhysRevC.70.034902}{\bibinfo {journal} {Phys.Rev.}}\ }%
  \textbf{\bibinfo {volume} {C70}},\ \bibinfo {pages} {034902} (\bibinfo {year}
  {2004})%
  \bibAnnoteFile{NoStop}{Anticic:2003fd}%
\bibitem{Adams:2004gp}%
  \BibitemOpen
  \bibfield{author}{%
  \bibinfo {author} {\bibfnamefont{J.}~\bibnamefont{Adams}} \emph{et~al.}
  (\bibinfo {collaboration} {STAR Collaboration}),\ }%
  \bibfield{journal}{%
  \Doi{10.1088/0954-3899/34/5/002}{\bibinfo {journal} {J.Phys.G}}\ }%
  \textbf{\bibinfo {volume} {G34}},\ \bibinfo {pages} {799} (\bibinfo {year}
  {2007})%
  \bibAnnoteFile{NoStop}{Adams:2004gp}%
\bibitem{Adams:2005ka}%
  \BibitemOpen
  \bibfield{author}{%
  \bibinfo {author} {\bibfnamefont{J.}~\bibnamefont{Adams}} \emph{et~al.}
  (\bibinfo {collaboration} {STAR Collaboration}),\ }%
  \bibfield{journal}{%
  \Doi{10.1103/PhysRevC.72.044902}{\bibinfo {journal} {Phys.Rev.}}\ }%
  \textbf{\bibinfo {volume} {C72}},\ \bibinfo {pages} {044902} (\bibinfo {year}
  {2005})%
  \bibAnnoteFile{NoStop}{Adams:2005ka}%
\bibitem{Adams:2005aw}%
  \BibitemOpen
  \bibfield{author}{%
  \bibinfo {author} {\bibfnamefont{J.}~\bibnamefont{Adams}} \emph{et~al.}
  (\bibinfo {collaboration} {STAR Collaboration}),\ }%
  \bibfield{journal}{%
  \Doi{10.1088/0954-3899/32/6/L02}{\bibinfo {journal} {J.Phys.G}}\ }%
  \textbf{\bibinfo {volume} {G32}},\ \bibinfo {pages} {L37} (\bibinfo {year}
  {2006})%
  \bibAnnoteFile{NoStop}{Adams:2005aw}%
\bibitem{Adams:2006sg}%
  \BibitemOpen
  \bibfield{author}{%
  \bibinfo {author} {\bibfnamefont{J.}~\bibnamefont{Adams}} \emph{et~al.}
  (\bibinfo {collaboration} {STAR Collaboration}),\ }%
  \bibfield{journal}{%
  \Doi{10.1088/0954-3899/34/3/004}{\bibinfo {journal} {J.Phys.G}}\ }%
  \textbf{\bibinfo {volume} {G34}},\ \bibinfo {pages} {451} (\bibinfo {year}
  {2007})%
  \bibAnnoteFile{NoStop}{Adams:2006sg}%
\bibitem{Grebieszkow:2007xz}%
  \BibitemOpen
  \bibfield{author}{%
  \bibinfo {author} {\bibfnamefont{K.}~\bibnamefont{Grebieszkow}}, \bibinfo
  {author} {\bibfnamefont{C.}~\bibnamefont{Alt}}, \bibinfo {author}
  {\bibfnamefont{T.}~\bibnamefont{Anticic}}, \bibinfo {author}
  {\bibfnamefont{B.}~\bibnamefont{Baatar}}, \bibinfo {author}
  {\bibfnamefont{D.}~\bibnamefont{Barna}}, \emph{et~al.},\ }%
  \bibfield{journal}{%
  \bibinfo {journal} {PoS}\ }%
  \textbf{\bibinfo {volume} {CPOD07}},\ \bibinfo {pages} {022} (\bibinfo {year}
  {2007})%
  \bibAnnoteFile{NoStop}{Grebieszkow:2007xz}%
\bibitem{na49:2008vb}%
  \BibitemOpen
  \bibfield{author}{%
  \bibinfo {author} {\bibfnamefont{T.}~\bibnamefont{Anticic}} \emph{et~al.}
  (\bibinfo {collaboration} {NA49}),\ }%
  \bibfield{journal}{%
  \bibinfo {journal} {Phys. Rev.}\ }%
  \textbf{\bibinfo {volume} {C79}},\ \bibinfo {pages} {044904} (\bibinfo {year}
  {2009})%
  \bibAnnoteFile{NoStop}{na49:2008vb}%
\bibitem{Adamova:2008sx}%
  \BibitemOpen
  \bibfield{author}{%
  \bibinfo {author} {\bibfnamefont{D.}~\bibnamefont{Adamova}} \emph{et~al.}
  (\bibinfo {collaboration} {CERES Collaboration}),\ }%
  \bibfield{journal}{%
  \Doi{10.1016/j.nuclphysa.2008.07.014}{\bibinfo {journal} {Nucl.Phys.}}\ }%
  \textbf{\bibinfo {volume} {A811}},\ \bibinfo {pages} {179} (\bibinfo {year}
  {2008})%
  \bibAnnoteFile{NoStop}{Adamova:2008sx}%
\bibitem{Agakishiev:2011fs}%
  \BibitemOpen
  \bibfield{author}{%
  \bibinfo {author} {\bibfnamefont{H.}~\bibnamefont{Agakishiev}} \emph{et~al.}
  (\bibinfo {collaboration} {STAR Collaboration}),\ }%
  \bibfield{journal}{%
  \Doi{10.1016/j.physletb.2011.09.075}{\bibinfo {journal} {Phys.Lett.}}\ }%
  \textbf{\bibinfo {volume} {B704}},\ \bibinfo {pages} {467} (\bibinfo {year}
  {2011})%
  \bibAnnoteFile{NoStop}{Agakishiev:2011fs}%
\bibitem{Bialas:1976ed}%
  \BibitemOpen
  \bibfield{author}{%
  \bibinfo {author} {\bibfnamefont{A.}~\bibnamefont{Bia\l{}as}}, \bibinfo
  {author} {\bibfnamefont{M.}~\bibnamefont{Bleszy\'nski}},\ and\ \bibinfo
  {author} {\bibfnamefont{W.}~\bibnamefont{Czy\.z}},\ }%
  \bibfield{journal}{%
  \bibinfo {journal} {Nucl. Phys.}\ }%
  \textbf{\bibinfo {volume} {B111}},\ \bibinfo {pages} {461} (\bibinfo {year}
  {1976})%
  \bibAnnoteFile{NoStop}{Bialas:1976ed}%
\bibitem{Bialas:2008zza}%
  \BibitemOpen
  \bibfield{author}{%
  \bibinfo {author} {\bibfnamefont{A.}~\bibnamefont{Bialas}},\ }%
  \bibfield{journal}{%
  \Doi{10.1088/0954-3899/35/4/044053}{\bibinfo {journal} {J. Phys.}}\ }%
  \textbf{\bibinfo {volume} {G35}},\ \bibinfo {pages} {044053} (\bibinfo {year}
  {2008})%
  \bibAnnoteFile{NoStop}{Bialas:2008zza}%
\bibitem{Kharzeev:2000ph}%
  \BibitemOpen
  \bibfield{author}{%
  \bibinfo {author} {\bibfnamefont{D.}~\bibnamefont{Kharzeev}}\ and\ \bibinfo
  {author} {\bibfnamefont{M.}~\bibnamefont{Nardi}},\ }%
  \bibfield{journal}{%
  \Doi{10.1016/S0370-2693(01)00457-9}{\bibinfo {journal} {Phys. Lett.}}\ }%
  \textbf{\bibinfo {volume} {B507}},\ \bibinfo {pages} {121} (\bibinfo {year}
  {2001})%
  \bibAnnoteFile{NoStop}{Kharzeev:2000ph}%
\bibitem{Aguiar:2000hw}%
  \BibitemOpen
  \bibfield{author}{%
  \bibinfo {author} {\bibfnamefont{C.~E.}\ \bibnamefont{Aguiar}}, \bibinfo
  {author} {\bibfnamefont{T.}~\bibnamefont{Kodama}}, \bibinfo {author}
  {\bibfnamefont{T.}~\bibnamefont{Osada}},\ and\ \bibinfo {author}
  {\bibfnamefont{Y.}~\bibnamefont{Hama}},\ }%
  \bibfield{journal}{%
  \bibinfo {journal} {J. Phys.}\ }%
  \textbf{\bibinfo {volume} {G27}},\ \bibinfo {pages} {75} (\bibinfo {year}
  {2001})%
  \bibAnnoteFile{NoStop}{Aguiar:2000hw}%
\bibitem{Miller:2003kd}%
  \BibitemOpen
  \bibfield{author}{%
  \bibinfo {author} {\bibfnamefont{M.}~\bibnamefont{Miller}}\ and\ \bibinfo
  {author} {\bibfnamefont{R.}~\bibnamefont{Snellings}}}%
   (\bibinfo {year} {2003}),\
  \Eprint{http://arxiv.org/abs/nucl-ex/0312008}{nucl-ex/0312008}%
  \bibAnnoteFile{NoStop}{Miller:2003kd}%
\bibitem{Bhalerao:2005mm}%
  \BibitemOpen
  \bibfield{author}{%
  \bibinfo {author} {\bibfnamefont{R.~S.}\ \bibnamefont{Bhalerao}}, \bibinfo
  {author} {\bibfnamefont{J.-P.}\ \bibnamefont{Blaizot}}, \bibinfo {author}
  {\bibfnamefont{N.}~\bibnamefont{Borghini}},\ and\ \bibinfo {author}
  {\bibfnamefont{J.-Y.}\ \bibnamefont{Ollitrault}},\ }%
  \bibfield{journal}{%
  \Doi{10.1016/j.physletb.2005.08.131}{\bibinfo {journal} {Phys. Lett.}}\ }%
  \textbf{\bibinfo {volume} {B627}},\ \bibinfo {pages} {49} (\bibinfo {year}
  {2005})%
  \bibAnnoteFile{NoStop}{Bhalerao:2005mm}%
\bibitem{Manly:2005zy}%
  \BibitemOpen
  \bibfield{author}{%
  \bibinfo {author} {\bibfnamefont{S.}~\bibnamefont{Manly}} \emph{et~al.}
  (\bibinfo {collaboration} {PHOBOS}),\ }%
  \bibfield{journal}{%
  \Doi{10.1016/j.nuclphysa.2006.06.079}{\bibinfo {journal} {Nucl. Phys.}}\ }%
  \textbf{\bibinfo {volume} {A774}},\ \bibinfo {pages} {523} (\bibinfo {year}
  {2006})%
  \bibAnnoteFile{NoStop}{Manly:2005zy}%
\bibitem{Alver:2006wh}%
  \BibitemOpen
  \bibfield{author}{%
  \bibinfo {author} {\bibfnamefont{B.}~\bibnamefont{Alver}} \emph{et~al.}
  (\bibinfo {collaboration} {PHOBOS}),\ }%
  \bibfield{journal}{%
  \Doi{10.1103/PhysRevLett.98.242302}{\bibinfo {journal} {Phys. Rev. Lett.}}\
  }%
  \textbf{\bibinfo {volume} {98}},\ \bibinfo {pages} {242302} (\bibinfo {year}
  {2007})%
  \bibAnnoteFile{NoStop}{Alver:2006wh}%
\bibitem{Andrade:2006yh}%
  \BibitemOpen
  \bibfield{author}{%
  \bibinfo {author} {\bibfnamefont{R.}~\bibnamefont{Andrade}}, \bibinfo
  {author} {\bibfnamefont{F.}~\bibnamefont{Grassi}}, \bibinfo {author}
  {\bibfnamefont{Y.}~\bibnamefont{Hama}}, \bibinfo {author}
  {\bibfnamefont{T.}~\bibnamefont{Kodama}},\ and\ \bibinfo {author}
  {\bibfnamefont{J.}~\bibnamefont{Socolowski}, \bibfnamefont{O.}},\ }%
  \bibfield{journal}{%
  \bibinfo {journal} {Phys. Rev. Lett.}\ }%
  \textbf{\bibinfo {volume} {97}},\ \bibinfo {pages} {202302} (\bibinfo {year}
  {2006})%
  \bibAnnoteFile{NoStop}{Andrade:2006yh}%
\bibitem{Voloshin:2006gz}%
  \BibitemOpen
  \bibfield{author}{%
  \bibinfo {author} {\bibfnamefont{S.~A.}\ \bibnamefont{Voloshin}}}%
   (\bibinfo {year} {2006}),\
  \Eprint{http://arxiv.org/abs/nucl-th/0606022}{arXiv:nucl-th/0606022}%
  \bibAnnoteFile{NoStop}{Voloshin:2006gz}%
\bibitem{Drescher:2006ca}%
  \BibitemOpen
  \bibfield{author}{%
  \bibinfo {author} {\bibfnamefont{H.~J.}\ \bibnamefont{Drescher}}\ and\
  \bibinfo {author} {\bibfnamefont{Y.}~\bibnamefont{Nara}},\ }%
  \bibfield{journal}{%
  \bibinfo {journal} {Phys. Rev.}\ }%
  \textbf{\bibinfo {volume} {C75}},\ \bibinfo {pages} {034905} (\bibinfo {year}
  {2007})%
  \bibAnnoteFile{NoStop}{Drescher:2006ca}%
\bibitem{Broniowski:2007ft}%
  \BibitemOpen
  \bibfield{author}{%
  \bibinfo {author} {\bibfnamefont{W.}~\bibnamefont{Broniowski}}, \bibinfo
  {author} {\bibfnamefont{P.}~\bibnamefont{Bo\.zek}},\ and\ \bibinfo {author}
  {\bibfnamefont{M.}~\bibnamefont{Rybczy\'nski}},\ }%
  \bibfield{journal}{%
  \bibinfo {journal} {Phys. Rev.}\ }%
  \textbf{\bibinfo {volume} {C76}},\ \bibinfo {pages} {054905} (\bibinfo {year}
  {2007})%
  \bibAnnoteFile{NoStop}{Broniowski:2007ft}%
\bibitem{Hama:2007dq}%
  \BibitemOpen
  \bibfield{author}{%
  \bibinfo {author} {\bibfnamefont{Y.}~\bibnamefont{Hama}}, \bibinfo {author}
  {\bibfnamefont{R.}~\bibnamefont{Peterson~G.Andrade}}, \bibinfo {author}
  {\bibfnamefont{F.}~\bibnamefont{Grassi}}, \bibinfo {author}
  {\bibfnamefont{W.-L.}\ \bibnamefont{Qian}}, \bibinfo {author}
  {\bibfnamefont{T.}~\bibnamefont{Osada}}, \emph{et~al.},\ }%
  \bibfield{journal}{%
  \Doi{10.1134/S106377880809010X}{\bibinfo {journal} {Phys.Atom.Nucl.}}\ }%
  \textbf{\bibinfo {volume} {71}},\ \bibinfo {pages} {1558} (\bibinfo {year}
  {2008})%
  \bibAnnoteFile{NoStop}{Hama:2007dq}%
\bibitem{Voloshin:2007pc}%
  \BibitemOpen
  \bibfield{author}{%
  \bibinfo {author} {\bibfnamefont{S.~A.}\ \bibnamefont{Voloshin}}, \bibinfo
  {author} {\bibfnamefont{A.~M.}\ \bibnamefont{Poskanzer}}, \bibinfo {author}
  {\bibfnamefont{A.}~\bibnamefont{Tang}},\ and\ \bibinfo {author}
  {\bibfnamefont{G.}~\bibnamefont{Wang}},\ }%
  \bibfield{journal}{%
  \Doi{10.1016/j.physletb.2007.11.043}{\bibinfo {journal} {Phys. Lett.}}\ }%
  \textbf{\bibinfo {volume} {B659}},\ \bibinfo {pages} {537} (\bibinfo {year}
  {2008})%
  \bibAnnoteFile{NoStop}{Voloshin:2007pc}%
\bibitem{Andrade:2008fa}%
  \BibitemOpen
  \bibfield{author}{%
  \bibinfo {author} {\bibfnamefont{R.~P.~G.}\ \bibnamefont{Andrade}}
  \emph{et~al.},\ }%
  \bibfield{journal}{%
  \bibinfo {journal} {Acta Phys. Polon.}\ }%
  \textbf{\bibinfo {volume} {B40}},\ \bibinfo {pages} {993} (\bibinfo {year}
  {2009})%
  \bibAnnoteFile{NoStop}{Andrade:2008fa}%
\bibitem{Alver:2010gr}%
  \BibitemOpen
  \bibfield{author}{%
  \bibinfo {author} {\bibfnamefont{B.}~\bibnamefont{Alver}}\ and\ \bibinfo
  {author} {\bibfnamefont{G.}~\bibnamefont{Roland}},\ }%
  \bibfield{journal}{%
  \Doi{10.1103/PhysRevC.81.054905}{\bibinfo {journal} {Phys. Rev.}}\ }%
  \textbf{\bibinfo {volume} {C81}},\ \bibinfo {pages} {054905} (\bibinfo {year}
  {2010})%
  \bibAnnoteFile{NoStop}{Alver:2010gr}%
\bibitem{Alver:2010dn}%
  \BibitemOpen
  \bibfield{author}{%
  \bibinfo {author} {\bibfnamefont{B.~H.}\ \bibnamefont{Alver}}, \bibinfo
  {author} {\bibfnamefont{C.}~\bibnamefont{Gombeaud}}, \bibinfo {author}
  {\bibfnamefont{M.}~\bibnamefont{Luzum}},\ and\ \bibinfo {author}
  {\bibfnamefont{J.-Y.}\ \bibnamefont{Ollitrault}},\ }%
  \bibfield{journal}{%
  \Doi{10.1103/PhysRevC.82.034913}{\bibinfo {journal} {Phys. Rev.}}\ }%
  \textbf{\bibinfo {volume} {C82}},\ \bibinfo {pages} {034913} (\bibinfo {year}
  {2010})%
  \bibAnnoteFile{NoStop}{Alver:2010dn}%
\bibitem{Petersen:2010cw}%
  \BibitemOpen
  \bibfield{author}{%
  \bibinfo {author} {\bibfnamefont{H.}~\bibnamefont{Petersen}}, \bibinfo
  {author} {\bibfnamefont{G.-Y.}\ \bibnamefont{Qin}}, \bibinfo {author}
  {\bibfnamefont{S.~A.}\ \bibnamefont{Bass}},\ and\ \bibinfo {author}
  {\bibfnamefont{B.}~\bibnamefont{Muller}},\ }%
  \bibfield{journal}{%
  \Doi{10.1103/PhysRevC.82.041901}{\bibinfo {journal} {Phys.Rev.}}\ }%
  \textbf{\bibinfo {volume} {C82}},\ \bibinfo {pages} {041901} (\bibinfo {year}
  {2010})%
  \bibAnnoteFile{NoStop}{Petersen:2010cw}%
\bibitem{Bozek:2010vz}%
  \BibitemOpen
  \bibfield{author}{%
  \bibinfo {author} {\bibfnamefont{P.}~\bibnamefont{Bo\.zek}}, \bibinfo
  {author} {\bibfnamefont{W.}~\bibnamefont{Broniowski}},\ and\ \bibinfo
  {author} {\bibfnamefont{J.}~\bibnamefont{Moreira}},\ }%
  \bibfield{journal}{%
  \Doi{10.1103/PhysRevC.83.034911}{\bibinfo {journal} {Phys.Rev.}}\ }%
  \textbf{\bibinfo {volume} {C83}},\ \bibinfo {pages} {034911} (\bibinfo {year}
  {2011})%
  \bibAnnoteFile{NoStop}{Bozek:2010vz}%
\bibitem{Teaney:2010vd}%
  \BibitemOpen
  \bibfield{author}{%
  \bibinfo {author} {\bibfnamefont{D.}~\bibnamefont{Teaney}}\ and\ \bibinfo
  {author} {\bibfnamefont{L.}~\bibnamefont{Yan}},\ }%
  \bibfield{journal}{%
  \Doi{10.1103/PhysRevC.83.064904}{\bibinfo {journal} {Phys. Rev.}}\ }%
  \textbf{\bibinfo {volume} {C83}},\ \bibinfo {pages} {064904} (\bibinfo {year}
  {2011})%
  \bibAnnoteFile{NoStop}{Teaney:2010vd}%
\bibitem{Gardim:2011qn}%
  \BibitemOpen
  \bibfield{author}{%
  \bibinfo {author} {\bibfnamefont{F.~G.}\ \bibnamefont{Gardim}}, \bibinfo
  {author} {\bibfnamefont{F.}~\bibnamefont{Grassi}}, \bibinfo {author}
  {\bibfnamefont{Y.}~\bibnamefont{Hama}}, \bibinfo {author}
  {\bibfnamefont{M.}~\bibnamefont{Luzum}},\ and\ \bibinfo {author}
  {\bibfnamefont{J.-Y.}\ \bibnamefont{Ollitrault}},\ }%
  \bibfield{journal}{%
  \Doi{10.1103/PhysRevC.83.064901}{\bibinfo {journal} {Phys. Rev.}}\ }%
  \textbf{\bibinfo {volume} {C83}},\ \bibinfo {pages} {064901} (\bibinfo {year}
  {2011})%
  \bibAnnoteFile{NoStop}{Gardim:2011qn}%
\bibitem{Back:2001xy}%
  \BibitemOpen
  \bibfield{author}{%
  \bibinfo {author} {\bibfnamefont{B.~B.}\ \bibnamefont{Back}} \emph{et~al.}
  (\bibinfo {collaboration} {PHOBOS}),\ }%
  \bibfield{journal}{%
  \bibinfo {journal} {Phys. Rev.}\ }%
  \textbf{\bibinfo {volume} {C65}},\ \bibinfo {pages} {031901} (\bibinfo {year}
  {2002})%
  \bibAnnoteFile{NoStop}{Back:2001xy}%
\bibitem{Back:2004dy}%
  \BibitemOpen
  \bibfield{author}{%
  \bibinfo {author} {\bibfnamefont{B.~B.}\ \bibnamefont{Back}} \emph{et~al.}
  (\bibinfo {collaboration} {PHOBOS}),\ }%
  \bibfield{journal}{%
  \bibinfo {journal} {Phys. Rev.}\ }%
  \textbf{\bibinfo {volume} {C70}},\ \bibinfo {pages} {021902} (\bibinfo {year}
  {2004})%
  \bibAnnoteFile{NoStop}{Back:2004dy}%
\bibitem{Rybczynski:2011wv}%
  \BibitemOpen
  \bibfield{author}{%
  \bibinfo {author} {\bibfnamefont{M.}~\bibnamefont{Rybczynski}}\ and\ \bibinfo
  {author} {\bibfnamefont{W.}~\bibnamefont{Broniowski}},\ }%
  \bibfield{journal}{%
  \Doi{10.1103/PhysRevC.84.064913}{\bibinfo {journal} {Phys.Rev.}}\ }%
  \textbf{\bibinfo {volume} {C84}},\ \bibinfo {pages} {064913} (\bibinfo {year}
  {2011}),\ \bibinfo {note} {8 pages, 7 figures}%
  \bibAnnoteFile{NoStop}{Rybczynski:2011wv}%
\bibitem{Bialas:2004su}%
  \BibitemOpen
  \bibfield{author}{%
  \bibinfo {author} {\bibfnamefont{A.}~\bibnamefont{Bia\l{}as}}\ and\ \bibinfo
  {author} {\bibfnamefont{W.}~\bibnamefont{Czy\.z}},\ }%
  \bibfield{journal}{%
  \bibinfo {journal} {Acta Phys. Polon.}\ }%
  \textbf{\bibinfo {volume} {B36}},\ \bibinfo {pages} {905} (\bibinfo {year}
  {2005})%
  \bibAnnoteFile{NoStop}{Bialas:2004su}%
\bibitem{Bozek:2010bi}%
  \BibitemOpen
  \bibfield{author}{%
  \bibinfo {author} {\bibfnamefont{P.}~\bibnamefont{Bo\.zek}}\ and\ \bibinfo
  {author} {\bibfnamefont{I.}~\bibnamefont{Wyskiel}},\ }%
  \bibfield{journal}{%
  \Doi{10.1103/PhysRevC.81.054902}{\bibinfo {journal} {Phys. Rev.}}\ }%
  \textbf{\bibinfo {volume} {C81}},\ \bibinfo {pages} {054902} (\bibinfo {year}
  {2010})%
  \bibAnnoteFile{NoStop}{Bozek:2010bi}%
\bibitem{Bozek:2010er}%
  \BibitemOpen
  \bibfield{author}{%
  \bibinfo {author} {\bibfnamefont{P.}~\bibnamefont{Bo\.zek}},\ }%
  \bibfield{journal}{%
  \Doi{10.1103/PhysRevC.83.044910}{\bibinfo {journal} {Phys. Rev.}}\ }%
  \textbf{\bibinfo {volume} {C83}},\ \bibinfo {pages} {044910} (\bibinfo {year}
  {2011})%
  \bibAnnoteFile{NoStop}{Bozek:2010er}%
\bibitem{Bozek:2011wa}%
  \BibitemOpen
  \bibfield{author}{%
  \bibinfo {author} {\bibfnamefont{P.}~\bibnamefont{Bo\.zek}},\ }%
  \bibfield{journal}{%
  \Doi{10.1016/j.physletb.2011.04.020}{\bibinfo {journal} {Phys. Lett.}}\ }%
  \textbf{\bibinfo {volume} {B699}},\ \bibinfo {pages} {283} (\bibinfo {year}
  {2011})%
  \bibAnnoteFile{NoStop}{Bozek:2011wa}%
\bibitem{Bialas:2006kw}%
  \BibitemOpen
  \bibfield{author}{%
  \bibinfo {author} {\bibfnamefont{A.}~\bibnamefont{Bia\l{}as}}\ and\ \bibinfo
  {author} {\bibfnamefont{A.}~\bibnamefont{Bzdak}},\ }%
  \bibfield{journal}{%
  \bibinfo {journal} {Phys. Lett.}\ }%
  \textbf{\bibinfo {volume} {B649}},\ \bibinfo {pages} {263} (\bibinfo {year}
  {2007})%
  \bibAnnoteFile{NoStop}{Bialas:2006kw}%
\bibitem{Alvioli:2009ab}%
  \BibitemOpen
  \bibfield{author}{%
  \bibinfo {author} {\bibfnamefont{M.}~\bibnamefont{Alvioli}}, \bibinfo
  {author} {\bibfnamefont{H.-J.}\ \bibnamefont{Drescher}},\ and\ \bibinfo
  {author} {\bibfnamefont{M.}~\bibnamefont{Strikman}},\ }%
  \bibfield{journal}{%
  \Doi{10.1016/j.physletb.2009.08.067}{\bibinfo {journal} {Phys.Lett.}}\ }%
  \textbf{\bibinfo {volume} {B680}},\ \bibinfo {pages} {225} (\bibinfo {year}
  {2009})%
  \bibAnnoteFile{NoStop}{Alvioli:2009ab}%
\bibitem{Broniowski:2010jd}%
  \BibitemOpen
  \bibfield{author}{%
  \bibinfo {author} {\bibfnamefont{W.}~\bibnamefont{Broniowski}}\ and\ \bibinfo
  {author} {\bibfnamefont{M.}~\bibnamefont{Rybczynski}},\ }%
  \bibfield{journal}{%
  \Doi{10.1103/PhysRevC.81.064909}{\bibinfo {journal} {Phys.Rev.}}\ }%
  \textbf{\bibinfo {volume} {C81}},\ \bibinfo {pages} {064909} (\bibinfo {year}
  {2010})%
  \bibAnnoteFile{NoStop}{Broniowski:2010jd}%
\bibitem{Alvioli:2010yk}%
  \BibitemOpen
  \bibfield{author}{%
  \bibinfo {author} {\bibfnamefont{M.}~\bibnamefont{Alvioli}}\ and\ \bibinfo
  {author} {\bibfnamefont{M.}~\bibnamefont{Strikman}},\ }%
  \bibfield{journal}{%
  \Doi{10.1103/PhysRevC.83.044905}{\bibinfo {journal} {Phys.Rev.}}\ }%
  \textbf{\bibinfo {volume} {C83}},\ \bibinfo {pages} {044905} (\bibinfo {year}
  {2011})%
  \bibAnnoteFile{NoStop}{Alvioli:2010yk}%
\bibitem{Kolb:2003dz}%
  \BibitemOpen
  \bibfield{author}{%
  \bibinfo {author} {\bibfnamefont{P.~F.}\ \bibnamefont{Kolb}}\ and\ \bibinfo
  {author} {\bibfnamefont{U.~W.}\ \bibnamefont{Heinz}},\ }%
  in\ \emph{\bibinfo {booktitle} {Quark Gluon Plasma 3}},\ \bibinfo {editor}
  {edited by\ \bibinfo {editor} {\bibfnamefont{R.}~\bibnamefont{Hwa}}\ and\
  \bibinfo {editor} {\bibfnamefont{X.~N.}\ \bibnamefont{Wang}}}\ (\bibinfo
  {publisher} {World Scientific, Singapore},\ \bibinfo {year} {2004})\ p.\
  \bibinfo {pages} {634},\
  \Eprint{http://arxiv.org/abs/nucl-th/0305084}{arXiv:nucl-th/0305084}%
  \bibAnnoteFile{NoStop}{Kolb:2003dz}%
\bibitem{Huovinen:2006jp}%
  \BibitemOpen
  \bibfield{author}{%
  \bibinfo {author} {\bibfnamefont{P.}~\bibnamefont{Huovinen}}\ and\ \bibinfo
  {author} {\bibfnamefont{P.~V.}\ \bibnamefont{Ruuskanen}},\ }%
  \bibfield{journal}{%
  \Doi{10.1146/annurev.nucl.54.070103.181236}{\bibinfo {journal} {Ann. Rev.
  Nucl. Part. Sci.}}\ }%
  \textbf{\bibinfo {volume} {56}},\ \bibinfo {pages} {163} (\bibinfo {year}
  {2006})%
  \bibAnnoteFile{NoStop}{Huovinen:2006jp}%
\bibitem{Florkowski:2010zz}%
  \BibitemOpen
  \bibfield{author}{%
  \bibinfo {author} {\bibfnamefont{W.}~\bibnamefont{Florkowski}},\ }%
  \emph{\bibinfo {title} {{Phenomenology of Ultra-Relativistic Heavy-Ion
  Collisions}}}\ (\bibinfo {publisher} {World Scientific Publishing Company,
  Singapore},\ \bibinfo {year} {2010})%
  \bibAnnoteFile{NoStop}{Florkowski:2010zz}%
\bibitem{Werner:2009fa}%
  \BibitemOpen
  \bibfield{author}{%
  \bibinfo {author} {\bibfnamefont{K.}~\bibnamefont{Werner}} \emph{et~al.},\ }%
  \bibfield{journal}{%
  \bibinfo {journal} {J. Phys.}\ }%
  \textbf{\bibinfo {volume} {G36}},\ \bibinfo {pages} {064030} (\bibinfo {year}
  {2009})%
  \bibAnnoteFile{NoStop}{Werner:2009fa}%
\bibitem{Holopainen:2010gz}%
  \BibitemOpen
  \bibfield{author}{%
  \bibinfo {author} {\bibfnamefont{H.}~\bibnamefont{Holopainen}}, \bibinfo
  {author} {\bibfnamefont{H.}~\bibnamefont{Niemi}},\ and\ \bibinfo {author}
  {\bibfnamefont{K.~J.}\ \bibnamefont{Eskola}},\ }%
  \bibfield{journal}{%
  \Doi{10.1103/PhysRevC.83.034901}{\bibinfo {journal} {Phys.Rev.}}\ }%
  \textbf{\bibinfo {volume} {C83}},\ \bibinfo {pages} {034901} (\bibinfo {year}
  {2011})%
  \bibAnnoteFile{NoStop}{Holopainen:2010gz}%
\bibitem{Gardim:2011xv}%
  \BibitemOpen
  \bibfield{author}{%
  \bibinfo {author} {\bibfnamefont{F.~G.}\ \bibnamefont{Gardim}}, \bibinfo
  {author} {\bibfnamefont{F.}~\bibnamefont{Grassi}}, \bibinfo {author}
  {\bibfnamefont{M.}~\bibnamefont{Luzum}},\ and\ \bibinfo {author}
  {\bibfnamefont{J.-Y.}\ \bibnamefont{Ollitrault}},\ }%
  \bibfield{journal}{%
  \Doi{10.1103/PhysRevC.85.024908}{\bibinfo {journal} {Phys. Rev.}}\ }%
  \textbf{\bibinfo {volume} {C85}},\ \bibinfo {pages} {024908} (\bibinfo {year}
  {2012})%
  \bibAnnoteFile{NoStop}{Gardim:2011xv}%
\bibitem{Schenke:2010rr}%
  \BibitemOpen
  \bibfield{author}{%
  \bibinfo {author} {\bibfnamefont{B.}~\bibnamefont{Schenke}}, \bibinfo
  {author} {\bibfnamefont{S.}~\bibnamefont{Jeon}},\ and\ \bibinfo {author}
  {\bibfnamefont{C.}~\bibnamefont{Gale}},\ }%
  \bibfield{journal}{%
  \Doi{10.1103/PhysRevLett.106.042301}{\bibinfo {journal} {Phys. Rev. Lett.}}\
  }%
  \textbf{\bibinfo {volume} {106}},\ \bibinfo {pages} {042301} (\bibinfo {year}
  {2011})%
  \bibAnnoteFile{NoStop}{Schenke:2010rr}%
\bibitem{Qiu:2011fi}%
  \BibitemOpen
  \bibfield{author}{%
  \bibinfo {author} {\bibfnamefont{Z.}~\bibnamefont{Qiu}}\ and\ \bibinfo
  {author} {\bibfnamefont{U.~W.}\ \bibnamefont{Heinz}}}%
   (\bibinfo {year} {2011}),\
  \Eprint{http://arxiv.org/abs/1108.1714}{arXiv:1108.1714 [nucl-th]}%
  \bibAnnoteFile{NoStop}{Qiu:2011fi}%
\bibitem{Chaudhuri:2011pa}%
  \BibitemOpen
  \bibfield{author}{%
  \bibinfo {author} {\bibfnamefont{A.}~\bibnamefont{Chaudhuri}}}%
   (\bibinfo {year} {2011}),\
  \Eprint{http://arxiv.org/abs/1112.1166}{arXiv:1112.1166 [nucl-th]}%
  \bibAnnoteFile{NoStop}{Chaudhuri:2011pa}%
\bibitem{IS}%
  \BibitemOpen
  \bibfield{author}{%
  \bibinfo {author} {\bibfnamefont{W.}~\bibnamefont{Israel}}\ and\ \bibinfo
  {author} {\bibfnamefont{J.}~\bibnamefont{Stewart}},\ }%
  \bibfield{journal}{%
  \bibinfo {journal} {Annals Phys.}\ }%
  \textbf{\bibinfo {volume} {118}},\ \bibinfo {pages} {341} (\bibinfo {year}
  {1979})%
  \bibAnnoteFile{NoStop}{IS}%
\bibitem{Romatschke:2009im}%
  \BibitemOpen
  \bibfield{author}{%
  \bibinfo {author} {\bibfnamefont{P.}~\bibnamefont{Romatschke}},\ }%
  \bibfield{journal}{%
  \Doi{10.1142/S0218301310014613}{\bibinfo {journal} {Int. J. Mod. Phys.}}\ }%
  \textbf{\bibinfo {volume} {E19}},\ \bibinfo {pages} {1} (\bibinfo {year}
  {2010})%
  \bibAnnoteFile{NoStop}{Romatschke:2009im}%
\bibitem{Teaney:2009qa}%
  \BibitemOpen
  \bibfield{author}{%
  \bibinfo {author} {\bibfnamefont{D.~A.}\ \bibnamefont{Teaney}}}%
   (\bibinfo {year} {2009}),\
  \Eprint{http://arxiv.org/abs/0905.2433}{arXiv:0905.2433 [nucl-th]}%
  \bibAnnoteFile{NoStop}{Teaney:2009qa}%
\bibitem{Borsanyi:2010cj}%
  \BibitemOpen
  \bibfield{author}{%
  \bibinfo {author} {\bibfnamefont{S.}~\bibnamefont{Borsanyi}} \emph{et~al.},\
  }%
  \bibfield{journal}{%
  \Doi{10.1007/JHEP11(2010)077}{\bibinfo {journal} {JHEP}}\ }%
  \textbf{\bibinfo {volume} {11}},\ \bibinfo {pages} {077} (\bibinfo {year}
  {2010})%
  \bibAnnoteFile{NoStop}{Borsanyi:2010cj}%
\bibitem{Chojnacki:2007jc}%
  \BibitemOpen
  \bibfield{author}{%
  \bibinfo {author} {\bibfnamefont{M.}~\bibnamefont{Chojnacki}}\ and\ \bibinfo
  {author} {\bibfnamefont{W.}~\bibnamefont{Florkowski}},\ }%
  \bibfield{journal}{%
  \bibinfo {journal} {Acta Phys. Polon.}\ }%
  \textbf{\bibinfo {volume} {B38}},\ \bibinfo {pages} {3249} (\bibinfo {year}
  {2007})%
  \bibAnnoteFile{NoStop}{Chojnacki:2007jc}%
\bibitem{Broniowski:2001ei}%
  \BibitemOpen
  \bibfield{author}{%
  \bibinfo {author} {\bibfnamefont{W.}~\bibnamefont{Broniowski}}\ and\ \bibinfo
  {author} {\bibfnamefont{W.}~\bibnamefont{Florkowski}},\ }%
  \bibfield{journal}{%
  \Doi{10.1103/PhysRevC.65.024905}{\bibinfo {journal} {Phys.Rev.}}\ }%
  \textbf{\bibinfo {volume} {C65}},\ \bibinfo {pages} {024905} (\bibinfo {year}
  {2002})%
  \bibAnnoteFile{NoStop}{Broniowski:2001ei}%
\bibitem{Broniowski:2002nf}%
  \BibitemOpen
  \bibfield{author}{%
  \bibinfo {author} {\bibfnamefont{W.}~\bibnamefont{Broniowski}}, \bibinfo
  {author} {\bibfnamefont{A.}~\bibnamefont{Baran}},\ and\ \bibinfo {author}
  {\bibfnamefont{W.}~\bibnamefont{Florkowski}},\ }%
  \bibfield{journal}{%
  \bibinfo {journal} {Acta Phys. Polon.}\ }%
  \textbf{\bibinfo {volume} {B33}},\ \bibinfo {pages} {4235} (\bibinfo {year}
  {2002})%
  \bibAnnoteFile{NoStop}{Broniowski:2002nf}%
\bibitem{Torrieri:2004zz}%
  \BibitemOpen
  \bibfield{author}{%
  \bibinfo {author} {\bibfnamefont{G.}~\bibnamefont{Torrieri}} \emph{et~al.},\
  }%
  \bibfield{journal}{%
  \bibinfo {journal} {Comput. Phys. Commun.}\ }%
  \textbf{\bibinfo {volume} {167}},\ \bibinfo {pages} {229} (\bibinfo {year}
  {2005})%
  \bibAnnoteFile{NoStop}{Torrieri:2004zz}%
\bibitem{Teaney:2003kp}%
  \BibitemOpen
  \bibfield{author}{%
  \bibinfo {author} {\bibfnamefont{D.}~\bibnamefont{Teaney}},\ }%
  \bibfield{journal}{%
  \bibinfo {journal} {Phys. Rev.}\ }%
  \textbf{\bibinfo {volume} {C68}},\ \bibinfo {pages} {034913} (\bibinfo {year}
  {2003})%
  \bibAnnoteFile{NoStop}{Teaney:2003kp}%
\bibitem{Gavin:1985ph}%
  \BibitemOpen
  \bibfield{author}{%
  \bibinfo {author} {\bibfnamefont{S.}~\bibnamefont{Gavin}},\ }%
  \bibfield{journal}{%
  \Doi{10.1016/0375-9474(85)90190-3}{\bibinfo {journal} {Nucl. Phys.}}\ }%
  \textbf{\bibinfo {volume} {A435}},\ \bibinfo {pages} {826} (\bibinfo {year}
  {1985})%
  \bibAnnoteFile{NoStop}{Gavin:1985ph}%
\bibitem{Hosoya:1983xm}%
  \BibitemOpen
  \bibfield{author}{%
  \bibinfo {author} {\bibfnamefont{A.}~\bibnamefont{Hosoya}}\ and\ \bibinfo
  {author} {\bibfnamefont{K.}~\bibnamefont{Kajantie}},\ }%
  \bibfield{journal}{%
  \Doi{10.1016/0550-3213(85)90499-7}{\bibinfo {journal} {Nucl. Phys.}}\ }%
  \textbf{\bibinfo {volume} {B250}},\ \bibinfo {pages} {666} (\bibinfo {year}
  {1985})%
  \bibAnnoteFile{NoStop}{Hosoya:1983xm}%
\bibitem{Sasaki:2008fg}%
  \BibitemOpen
  \bibfield{author}{%
  \bibinfo {author} {\bibfnamefont{C.}~\bibnamefont{Sasaki}}\ and\ \bibinfo
  {author} {\bibfnamefont{K.}~\bibnamefont{Redlich}},\ }%
  \bibfield{journal}{%
  \Doi{10.1103/PhysRevC.79.055207}{\bibinfo {journal} {Phys. Rev.}}\ }%
  \textbf{\bibinfo {volume} {C79}},\ \bibinfo {pages} {055207} (\bibinfo {year}
  {2009})%
  \bibAnnoteFile{NoStop}{Sasaki:2008fg}%
\bibitem{Bozek:2009dw}%
  \BibitemOpen
  \bibfield{author}{%
  \bibinfo {author} {\bibfnamefont{P.}~\bibnamefont{Bo\.zek}},\ }%
  \bibfield{journal}{%
  \bibinfo {journal} {Phys. Rev.}\ }%
  \textbf{\bibinfo {volume} {C81}},\ \bibinfo {pages} {034909} (\bibinfo {year}
  {2010})%
  \bibAnnoteFile{NoStop}{Bozek:2009dw}%
\bibitem{Andronic:2005yp}%
  \BibitemOpen
  \bibfield{author}{%
  \bibinfo {author} {\bibfnamefont{A.}~\bibnamefont{Andronic}}, \bibinfo
  {author} {\bibfnamefont{P.}~\bibnamefont{Braun-Munzinger}},\ and\ \bibinfo
  {author} {\bibfnamefont{J.}~\bibnamefont{Stachel}},\ }%
  \bibfield{journal}{%
  \Doi{10.1016/j.nuclphysa.2006.03.012}{\bibinfo {journal} {Nucl. Phys.}}\ }%
  \textbf{\bibinfo {volume} {A772}},\ \bibinfo {pages} {167} (\bibinfo {year}
  {2006})%
  \bibAnnoteFile{NoStop}{Andronic:2005yp}%
\bibitem{Adler:2003cb}%
  \BibitemOpen
  \bibfield{author}{%
  \bibinfo {author} {\bibfnamefont{S.~S.}\ \bibnamefont{Adler}} \emph{et~al.}
  (\bibinfo {collaboration} {PHENIX}),\ }%
  \bibfield{journal}{%
  \Doi{10.1103/PhysRevC.69.034909}{\bibinfo {journal} {Phys. Rev.}}\ }%
  \textbf{\bibinfo {volume} {C69}},\ \bibinfo {pages} {034909} (\bibinfo {year}
  {2004})%
  \bibAnnoteFile{NoStop}{Adler:2003cb}%
\bibitem{Andrade:2008xh}%
  \BibitemOpen
  \bibfield{author}{%
  \bibinfo {author} {\bibfnamefont{R.}~\bibnamefont{Andrade}}, \bibinfo
  {author} {\bibfnamefont{F.}~\bibnamefont{Grassi}}, \bibinfo {author}
  {\bibfnamefont{Y.}~\bibnamefont{Hama}}, \bibinfo {author}
  {\bibfnamefont{T.}~\bibnamefont{Kodama}},\ and\ \bibinfo {author}
  {\bibfnamefont{W.}~\bibnamefont{Qian}},\ }%
  \bibfield{journal}{%
  \Doi{10.1103/PhysRevLett.101.112301}{\bibinfo {journal} {Phys.Rev.Lett.}}\ }%
  \textbf{\bibinfo {volume} {101}},\ \bibinfo {pages} {112301} (\bibinfo {year}
  {2008})%
  \bibAnnoteFile{NoStop}{Andrade:2008xh}%
\bibitem{Ollitrault:1991xx}%
  \BibitemOpen
  \bibfield{author}{%
  \bibinfo {author} {\bibfnamefont{J.-Y.}\ \bibnamefont{Ollitrault}},\ }%
  \bibfield{journal}{%
  \Doi{10.1016/0370-2693(91)90548-5}{\bibinfo {journal} {Phys.Lett.}}\ }%
  \textbf{\bibinfo {volume} {B273}},\ \bibinfo {pages} {32} (\bibinfo {year}
  {1991})%
  \bibAnnoteFile{NoStop}{Ollitrault:1991xx}%
\bibitem{Broniowski:2009te}%
  \BibitemOpen
  \bibfield{author}{%
  \bibinfo {author} {\bibfnamefont{W.}~\bibnamefont{Broniowski}}, \bibinfo
  {author} {\bibfnamefont{M.}~\bibnamefont{Rybczynski}}, \bibinfo {author}
  {\bibfnamefont{L.}~\bibnamefont{Obara}},\ and\ \bibinfo {author}
  {\bibfnamefont{M.}~\bibnamefont{Chojnacki}},\ }%
  \bibfield{journal}{%
  \bibinfo {journal} {Acta Phys.Polon.Supp.}\ }%
  \textbf{\bibinfo {volume} {3}},\ \bibinfo {pages} {513} (\bibinfo {year}
  {2010})%
  \bibAnnoteFile{NoStop}{Broniowski:2009te}%
\bibitem{Broniowski:2006zz}%
  \BibitemOpen
  \bibfield{author}{%
  \bibinfo {author} {\bibfnamefont{W.}~\bibnamefont{Broniowski}}, \bibinfo
  {author} {\bibfnamefont{P.}~\bibnamefont{Bo\.zek}}, \bibinfo {author}
  {\bibfnamefont{W.}~\bibnamefont{Florkowski}},\ and\ \bibinfo {author}
  {\bibfnamefont{B.}~\bibnamefont{Hiller}},\ }%
  \bibfield{journal}{%
  \bibinfo {journal} {PoS}\ }%
  \textbf{\bibinfo {volume} {CFRNC2006}},\ \bibinfo {pages} {020} (\bibinfo
  {year} {2006})%
  \bibAnnoteFile{NoStop}{Broniowski:2006zz}%
\bibitem{Retinskaya:2012ky}%
  \BibitemOpen
  \bibfield{author}{%
  \bibinfo {author} {\bibfnamefont{E.}~\bibnamefont{Retinskaya}}, \bibinfo
  {author} {\bibfnamefont{M.}~\bibnamefont{Luzum}},\ and\ \bibinfo {author}
  {\bibfnamefont{J.-Y.}\ \bibnamefont{Ollitrault}}}%
   (\bibinfo {year} {2012}),\
  \Eprint{http://arxiv.org/abs/1203.0931}{arXiv:1203.0931 [nucl-th]}%
  \bibAnnoteFile{NoStop}{Retinskaya:2012ky}%
\bibitem{Hohne:2006ks}%
  \BibitemOpen
  \bibfield{author}{%
  \bibinfo {author} {\bibfnamefont{C.}~\bibnamefont{Hohne}}, \bibinfo {author}
  {\bibfnamefont{F.}~\bibnamefont{Puhlhofer}},\ and\ \bibinfo {author}
  {\bibfnamefont{R.}~\bibnamefont{Stock}},\ }%
  \bibfield{journal}{%
  \Doi{10.1016/j.physletb.2008.02.042}{\bibinfo {journal} {Phys. Lett.}}\ }%
  \textbf{\bibinfo {volume} {B640}},\ \bibinfo {pages} {96} (\bibinfo {year}
  {2006})%
  \bibAnnoteFile{NoStop}{Hohne:2006ks}%
\bibitem{Becattini:2008ya}%
  \BibitemOpen
  \bibfield{author}{%
  \bibinfo {author} {\bibfnamefont{F.}~\bibnamefont{Becattini}}\ and\ \bibinfo
  {author} {\bibfnamefont{J.}~\bibnamefont{Manninen}},\ }%
  \bibfield{journal}{%
  \Doi{10.1016/j.physletb.2009.01.066}{\bibinfo {journal} {Phys. Lett.}}\ }%
  \textbf{\bibinfo {volume} {B673}},\ \bibinfo {pages} {19} (\bibinfo {year}
  {2009})%
  \bibAnnoteFile{NoStop}{Becattini:2008ya}%
\bibitem{Bozek:2005eu}%
  \BibitemOpen
  \bibfield{author}{%
  \bibinfo {author} {\bibfnamefont{P.}~\bibnamefont{Bo\.zek}},\ }%
  \bibfield{journal}{%
  \bibinfo {journal} {Acta Phys. Polon.}\ }%
  \textbf{\bibinfo {volume} {B36}},\ \bibinfo {pages} {3071} (\bibinfo {year}
  {2005})%
  \bibAnnoteFile{NoStop}{Bozek:2005eu}%
\bibitem{Werner:2007bf}%
  \BibitemOpen
  \bibfield{author}{%
  \bibinfo {author} {\bibfnamefont{K.}~\bibnamefont{Werner}},\ }%
  \bibfield{journal}{%
  \Doi{10.1103/PhysRevLett.98.152301}{\bibinfo {journal} {Phys. Rev. Lett.}}\
  }%
  \textbf{\bibinfo {volume} {98}},\ \bibinfo {pages} {152301} (\bibinfo {year}
  {2007})%
  \bibAnnoteFile{NoStop}{Werner:2007bf}%
\bibitem{Bozek:2008zw}%
  \BibitemOpen
  \bibfield{author}{%
  \bibinfo {author} {\bibfnamefont{P.}~\bibnamefont{Bo\.zek}},\ }%
  \bibfield{journal}{%
  \Doi{10.1103/PhysRevC.79.054901}{\bibinfo {journal} {Phys. Rev.}}\ }%
  \textbf{\bibinfo {volume} {C79}},\ \bibinfo {pages} {054901} (\bibinfo {year}
  {2009})%
  \bibAnnoteFile{NoStop}{Bozek:2008zw}%
\bibitem{Florchinger:2011qf}%
  \BibitemOpen
  \bibfield{author}{%
  \bibinfo {author} {\bibfnamefont{S.}~\bibnamefont{Florchinger}}\ and\
  \bibinfo {author} {\bibfnamefont{U.~A.}\ \bibnamefont{Wiedemann}},\ }%
  \bibfield{journal}{%
  \Doi{10.1007/JHEP11(2011)100}{\bibinfo {journal} {JHEP}}\ }%
  \textbf{\bibinfo {volume} {11}},\ \bibinfo {pages} {100} (\bibinfo {year}
  {2011})%
  \bibAnnoteFile{NoStop}{Florchinger:2011qf}%
\bibitem{Kapusta:2012gm}%
  \BibitemOpen
  \bibfield{author}{%
  \bibinfo {author} {\bibfnamefont{J.}~\bibnamefont{Kapusta}}, \bibinfo
  {author} {\bibfnamefont{B.}~\bibnamefont{Mueller}},\ and\ \bibinfo {author}
  {\bibfnamefont{M.}~\bibnamefont{Stephanov}}}%
   (\bibinfo {year} {2012}),\
  \Eprint{http://arxiv.org/abs/1201.3405}{arXiv:1201.3405 [nucl-th]}%
  \bibAnnoteFile{NoStop}{Kapusta:2012gm}%
\end{thebibliography}%

\end{document}